\documentclass[letterpaper,11pt]{article}

\usepackage{amsmath}
\usepackage{amsfonts}
\usepackage{amssymb}
\usepackage{graphicx}
\usepackage{color}

\usepackage{graphicx}
\usepackage[body={6.5in, 9in}, right=1in, top=1in]{geometry}
\usepackage{amssymb}
\usepackage{amsmath} 
\usepackage[numbers,sort&compress]{natbib}

\usepackage{amstext}
\usepackage{epsfig}
\usepackage{verbatim} 
\usepackage{fancybox}
\usepackage{ulem}
\usepackage{enumitem}
\usepackage{subfigure}
\usepackage{bbm}
\usepackage{parskip}
\usepackage{dsfont}
\usepackage[numbers,sort&compress]{natbib}
\usepackage{cancel}
\usepackage{pifont}
\newcommand\ea{\end{align}}
\newcommand\ba{\begin{align}}
\newcommand\ee{\end{equation}}
\newcommand\be{\begin{equation}}
\newcommand\eea{\end{eqnarray}}
\newcommand\bea{\begin{eqnarray}}
\def\beq{\begin{eqnarray}}
\def\eeq{\end{eqnarray}}
\newcommand{\sfrac}[2]{{\textstyle\frac{#1}{#2}}}
\newcommand\pd{\partial}
\newcommand\mpl{M_{\rm Pl}}

\newcommand\na{\nabla}

\newcommand\la{\langle}
\newcommand\ra{\rangle}
\newcommand\ep{\epsilon}

\def\gsim{ \lower .75ex \hbox{$\sim$} \llap{\raise .27ex \hbox{$>$}} }
\def\lsim{ \lower .75ex \hbox{$\sim$} \llap{\raise .27ex \hbox{$<$}} }

\newcommand{\Comment}[1]{{}}
\definecolor{darkblue}{rgb}{0.15,0.35,0.55}
\definecolor{reddish}{rgb}{0.65, 0.2, 0.2}
\definecolor{darkred}{rgb}{0.7,0.3,0.3}
\definecolor{darkgreen}{rgb}{0.2,0.7,0.3}
\definecolor{darkblue2}{rgb}{0.3,0.4,0.9}
\definecolor{greyish}{rgb}{.8,.8,.8}
\usepackage[linktocpage=true]{hyperref}
\hypersetup{
colorlinks=true,
citecolor=darkblue,
linkcolor=reddish,
urlcolor=darkblue,
pdfauthor={},
pdftitle={},
pdfsubject={}
}

\usepackage{xcolor,colortbl}

\begin{document}
\def\thefootnote{\fnsymbol{footnote}}

\begin{center}
\LARGE{\textbf{Non-Trivial Checks of Novel Consistency Relations}} \\[0.5cm]
 
\large{Lasha  Berezhiani$^{\rm a}$, Justin Khoury$^{\rm a}$ and Junpu Wang$^{\rm b}$}
\\[0.5cm]

\small{
\textit{$^{\rm a}$ Center for Particle Cosmology, Department of Physics and Astronomy, \\ University of Pennsylvania, Philadelphia, PA 19104}}

\vspace{.2cm}

\small{
\textit{$^{\rm b}$ Department of Physics and Astronomy,\\
Johns Hopkins University, Baltimore, MD 21218}}

\vspace{.2cm}

\end{center}

\vspace{.6cm}

\hrule \vspace{0.2cm}
\centerline{\small{\bf Abstract}}
\vspace{-0.2cm}
{\small Single-field perturbations satisfy an infinite number of consistency relations constraining the squeezed limit of  
correlation functions at each order in the soft momentum. These can be understood as Ward identities for an infinite set of residual global
symmetries, or equivalently as Slavnov-Taylor identities for spatial diffeomorphisms. In this paper, we perform a number of novel, non-trivial
checks of the identities in the context of slow-roll single field inflationary models with arbitrary sound speed. We focus for concreteness on identities
involving 3-point functions with a soft external mode, and consider all possible scalar and tensor combinations for the hard-momentum modes. In all these cases,
we check the consistency relations up to and including cubic order in the soft momentum. For this purpose, we compute for the first time the 3-point functions
involving 2 scalars and 1 tensor, as well as 2 tensors and 1 scalar, for arbitrary sound speed.} 
\vspace{0.3cm}
\noindent
\hrule
\def\thefootnote{\arabic{footnote}}
\setcounter{footnote}{0}

\section{Introduction}

The consistency relations for adiabatic modes~\cite{Maldacena:2002vr,Creminelli:2004yq,Cheung:2007sv,Creminelli:2012ed,Hinterbichler:2012nm,Senatore:2012wy,Hinterbichler:2013dpa,Goldberger:2013rsa,Berezhiani:2013ewa,Pimentel:2013gza} serve as a powerful discriminating principle among various classes of inflationary models. For perturbations on a spatially-flat, Friedmann-Robertson-Walker (FRW) background, the consistency relations take the schematic form
\be
\lim_{\vec{q}\rightarrow 0} M_n {\partial^n \over \partial q^n}
\left( {1\over P_\zeta (q)} \langle \zeta(\vec q) {\cal O}(\vec{k}_1,\ldots,\vec{k}_N)\rangle + 
{1\over P_\gamma (q)} \langle \gamma(\vec q) {\cal O}(\vec{k}_1,\ldots,\vec{k}_N)\rangle\right)
\sim - M_n {\partial^n \over \partial k^n}  \langle {\cal O}(\vec{k}_1,\ldots,\vec{k}_N)
  \rangle \, ,
\label{schematic}
\ee
where $M_n$ is a suitable projector (with indices suppressed), and ${\cal O}(\vec{p}_1,\ldots,\vec{p}_N)$ represents an arbitrary equal-time product of scalar $\zeta$ and tensor $\gamma_{ij}$ perturbations, with momenta $\vec k_1, \ldots, \vec k_N$. At each integer order $n$, these identities constrain --- completely for $n=0,1$, and partially for $n\geq 2$ --- the $q^n$ behavior of an $N+1$-point correlation function with a soft scalar or
tensor mode in terms of an $N$-point function.  These relations are powerful probes of the inflationary era, since they hold in all models of `single-clock' inflation
in which the background is an attractor. Conversely, they can be violated if multiple fields contribute to density perturbations and/or $\zeta$ grows outside the
horizon~\cite{Cai:2009fn, Khoury:2008wj,Namjoo:2012aa,Chen:2013aj,Chen:2013eea}. Observationally, the consistency relations can be tested with the cosmic microwave background, as well as with the large scale structure~\cite{Kehagias:2013yd,Peloso:2013zw,Creminelli:2013mca,Creminelli:2013poa,Creminelli:2013nua,Peloso:2013spa,Kehagias:2013rpa,Valageas:2013cma}.

Just like the soft-pion theorems of the strong interactions, the cosmological consistency relations are the consequence of Ward identities
for spontaneously broken symmetries~\cite{Hinterbichler:2013dpa,Goldberger:2013rsa,Assassi:2012zq}. The symmetries in this case are
global, gauge-preserving spatial coordinate transformations. They map field configurations which fall off at infinity into those which do not.
Nevertheless, certain linear combinations of these transformations can be smoothly extended to physical configurations which fall off at
infinity, and as such constitute adiabatic modes~\cite{Weinberg:2003sw}.  

Recently, it has been shown that the consistency relations~(\ref{schematic}) all derive from a single, {\it master identity}, which follows from the Slavnov-Taylor identity for
spatial diffeomorphisms~\cite{Berezhiani:2013ewa}. (See~\cite{Pimentel:2013gza} for an alternative derivation using the wavefunction.) The approach underscores
that the consistency conditions are the consequence of the underlying diffeomorphism invariance, despite the fact that the general coordinate invariance is not the 
symmetry of the action upon introduction of the gauge-fixing term for the graviton. The master identity is valid at any value of $q$ and therefore goes beyond the soft limit.
By differentiating it $n$ times with respect to $q$ and translating to correlation functions, one recovers~(\ref{schematic}) at each order~\cite{Berezhiani:2013ewa}. 

For instance, the identity for soft 3-point vertices with hard scalar modes takes the form~\cite{Berezhiani:2013ewa}
\beq
q^j\left(\frac{1}{3}\Gamma^{\zeta\zeta\zeta}(\vec{q},\vec{k},-\vec{q}-\vec{k})+2\Gamma_{ij}^{\gamma\zeta\zeta}(\vec{q},\vec{k},-\vec{q}-\vec{k})\right)=q_i \Gamma_\zeta( k )-k_i\bigg( \Gamma_\zeta(|\vec{q}+\vec{k}|)-\Gamma_\zeta( k ) \bigg)\,,
\label{wardfintro}
\eeq
where $\Gamma^{\zeta\zeta\zeta}$ and $\Gamma_{ij}^{\gamma\zeta\zeta}$ are respectively the cubic vertex functions for 3 scalars, and for 2 scalars$-$1 tensor, each without the momentum-conserving delta function, while $\Gamma_\zeta$ is the inverse scalar propagator. The solution for the vertex functions can be obtained as a power series around $q = 0$, up to an arbitrary symmetric, transverse matrix $A_{ij}$. This arbitrary term is model-dependent, and hence contains physical information about the underlying theory. It stems from the fact that~(\ref{wardfintro}) only constrains the longitudinal components of the vertex functions. The key assumption underlying the consistency relations is that $A_{ij}$ is analytic in $q$, specifically that it starts at ${\cal O}(q^2)$. For standard inflationary scenarios, this is equivalent to the usual assumption of constant asymptotic solutions for the mode functions. For more exotic examples, such as khronon inflation~\cite{Creminelli:2012xb}, the analyticity criterion is the unambiguous one.

In this paper, we perform a number of novel, non-trivial checks of the identities~(\ref{schematic}) in the context of slow-roll single field inflationary models with arbitrary sound speed $c_s \neq 1$.
The lowest-order identities, $n=0$ and $n=1$, have been checked with various examples elsewhere~\cite{Creminelli:2012ed}. Our primary interest lies in the higher-order ($n\geq 2$)
identities.\footnote{Some simple checks of $n= 2$ and $n=3$ identities were performed in~\cite{Hinterbichler:2013dpa}.} For concreteness, we focus on soft 3-point function relations, involving 
as hard modes all possible combinations of $\zeta$'s and $\gamma$'s. In all these cases, we check the consistency relations up to and including $n=3$. For this purpose, the
correlation functions $\langle \zeta \gamma\zeta\rangle$ and $\langle \gamma \zeta\gamma\rangle$ with $c_s \neq 1$ are computed here for the first time, using the
techonology of the effective field theory of inflation~\cite{Cheung:2007st}.

Aside from the obvious upshot of establishing the validity of the identities, we are motivated by two observations:

\begin{enumerate}

\item At order $q^2$ and higher, only part of the correlation functions are constrained by the identities, while the remainder represents a model-dependent ({\it i.e.}, physical) piece.
For instance, the $q^2$ contribution in $\langle \zeta\zeta\zeta\rangle$ corresponds to spatial curvature~\cite{Creminelli:2013cga}. Therefore, when checking the identities, the projector $M_n$ in~(\ref{schematic}) plays a crucial role in extracting the relevant part of the correlators.

\item To date, no background-wave argument has been formulated for the $n\geq 2$ identities. In particular, at order $q^3$ the mode function of the long mode includes an imaginary piece,
\be
\zeta(q,\tau) = \frac{H}{\sqrt{4\epsilon M_{\rm Pl}^2 q^3} }(1-i q\tau) e^{iq\tau}\simeq  \frac{H}{\sqrt{4\epsilon M_{\rm Pl}^2 q^3}}\left ( 1 + \frac{q^2}{2} + i\frac{q^3}{3} + \ldots\right)\,,
\ee
and hence cannot be treated as completely classical.\footnote{We thank Paolo Creminelli and Leonardo Senatore for discussions on this point.} The fact that the consistency relation holds at this order, as verified,
tells us that at least part of the long mode is classical and can be removed by a suitable coordinate transformation.

\end{enumerate}

From a technical point of view, the checks performed are highly non-trivial because of the parametric dependence. Let us focus, for instance, on the hard modes being scalars, {\it i.e.}, $\zeta\zeta$.
In this case, the right-hand side of the identity~(\ref{schematic}) is proportional to  
\be
P_\zeta \sim \frac{1}{c_s \epsilon}\,.
\label{Pparam}
\ee
On the left-hand side of~(\ref{schematic}), meanwhile, the 3-point functions schematically have the following parametric dependence, to leading order in slow-roll parameters:
\bea
\nonumber
\frac{\langle \zeta\zeta\zeta\rangle}{P_\zeta} &\sim& \frac{1-c_s^2}{\epsilon c_s^3}q^2 + \ldots\,;\\ 
\frac{\langle \gamma\zeta\zeta\rangle}{P_\gamma} &\sim& \frac{1}{\epsilon c_s} \left( 1 +A q + B\frac{q^2}{c_s^2} + Cq^2 \ldots\right)  \,,
\label{3ptparam}
\eea
where $A$, $B$ and $C$ are constants. We make the following observations:

\begin{itemize}

\item At lowest order in $q$, namely $n=0$ and $n=1$, the identity~(\ref{schematic}) constrains each 3-point function separately
in terms of $P_\zeta$. Because $\langle\zeta\zeta\zeta\rangle$ starts at order $q^2$, both sides of the $n=0$ and $n=1$ scalar
relations must vanish identically, which is indeed the case~\cite{Creminelli:2012ed}. The soft tensor relations, on the other hand, are non-trivial.
As can be seen the parametric dependence of $\langle \gamma\zeta\zeta\rangle$ at order $q^0$ and $q$ matches that of $P_\zeta$,
and the consistency relations are satisfied.  

\item For $n\geq 2$, however, the 3-point correlators on the left-hand side of~(\ref{3ptparam}) contain terms proportional to $1/c_s^3$, which cannot be matched by $P_\zeta \sim 1/c_s$ on the right-hand side.
However, the identity~(\ref{schematic}) only constrains a particular linear combination of the 3-point correlators, and we will find indeed that the $1/c_s^3$ cancel out,
leaving a $1/c_s$ remainder that matches the right-hand side. 

\item In the limit $c_s \rightarrow 1$, the $\langle\zeta\zeta\zeta\rangle$ correlator vanishes ({\it i.e.}, becomes subleading in slow-roll parameters). As a result, to leading order in $\epsilon$, the identity~(\ref{schematic}) must be satisfied solely due to $\langle \gamma\zeta\zeta\rangle$. Our explicit calculations confirm this expectation.

\end{itemize}

The paper is organized as follows. We first derive the non-linear symmetries govering cosmological perturbations in $\zeta$-gauge (Sec.~\ref{symreview})
and briefly review the derivation of the corresponding Ward identities (Sec.~\ref{derivereview}). In Sec.~\ref{3ptcalc} we calculate the various 3-point functions
for arbitary sound speed $c_s = {\rm constant}$, to leading order in slow-roll parameters. We then turn to explicit checks of the Ward identities up to and including $q^3$ order,
with $\zeta\zeta$ (Sec.~\ref{zetazeta}), $\zeta\gamma$ (Sec.~\ref{zetagamma}) and $\gamma\gamma$ (Sec.~\ref{gammagamma}) as hard mode insertions. We summarize our results
in Sec.~\ref{conclusec}.

\section{Gauge-Preserving Coordinate Transformations}
\label{symreview}

In $\zeta$-gauge, the scalar field is unperturbed, $\phi = \bar{\phi}(t)$, and the spatial metric takes the form
\be\label{zeta gij}
h_{ij}=a^2(t) e^{2\zeta} \left(e^{\gamma}\right)_{ij}\,;\qquad \gamma^i_{~i} = 0\,;~~\partial^i \gamma_{ij} = 0\,.
\ee
Scalar perturbations are captured by the conformal mode $\zeta$; tensor modes are encoded in $\gamma_{ij}$.
This choice completely fixes the gauge, at least for diffeomorphisms that fall off sufficiently fast at spatial infinity.
However, there is an infinite number of residual, global coordinate transformations that diverge
at infinity~\cite{Hinterbichler:2012nm,Hinterbichler:2013dpa}. These hold on {\it any} spatially-flat
FRW background and do not rely on any slow-roll or quasi-de Sitter approximation.

Let us briefly review the derivation of these residual symmetries. Consider a general time-dependent
spatial diffeormorphism, $x^i \to x^i-\xi^i(t, \vec x)$. Being purely spatial, this transformation clearly leaves
the choice $\phi = \phi(t)$ invariant. The question is: under what conditions does it also leave~(\ref{zeta gij}) invariant?

At linear order in $\xi$, the spatial metric \eqref{zeta gij} transforms as 
\be
h'_{ij}=a^2 e^{2\zeta'} \left(e^{\gamma'}\right)_{ij}
=h_{ij}+\pd_i\xi^k h_{kj}
+\pd_j\xi^k h_{ki}
+\xi^k \pd_k h_{ij}+{\cal O}(\xi^2)\;,
\ee
or, explicitly, 
\be\label{zeta pres trans}
 e^{2\delta\zeta} \left(e^{\gamma+\delta \gamma}\right)_{ij}
=\left(1+2\xi^k\pd_k \zeta\right)\left(e^{\gamma}\right)_{ij}
+\pd_i\xi^k \left(e^{\gamma}\right)_{kj}
+\pd_j\xi^k \left(e^{\gamma}\right)_{ki}
+\xi^k\pd_k\left(e^\gamma\right)_{ij}\;,
\ee
where we have defined $\delta\zeta=\zeta'-\zeta$ and $\delta\gamma=\gamma'-\gamma$. Notice that even though $\zeta$ and $\gamma$ are not small, their variation $\delta \zeta, \delta \gamma \sim {\cal O}(\xi)$ is small.\footnote{Throughout this calculation we keep terms up to order ${\cal O}(\xi)$.} Multiplying both sides from the right by $\left(e^{-\gamma}\right)_{jm}$ gives
\be
 e^{2\delta\zeta} \left(e^{\gamma+\delta \gamma}\right)_{ij}\left(e^{-\gamma}\right)_{jm}
=(1+2\xi^k\pd_k \zeta)\delta_{im}
+\pd_i\xi^m
+\left(e^{-\gamma}\right)_{m j}\pd_j\xi^k \left(e^{\gamma}\right)_{ki}
+\xi^k\pd_k\left(e^\gamma\right)_{ij}\left(e^{-\gamma}\right)_{jm}\;.
\label{identity1}
\ee

We can solve for $\delta\zeta$ by tracing with $\delta^{ij}$ and applying the Hadamard lemma
\bea
\nonumber
e^{X+\delta X}e^{-X}&=&\mathbf{1}+\delta X +\frac{1}{2}[X,\delta X]+\frac{1}{3!}[X, \,[X,\,\delta X]]+\dots\nonumber\\
\left(\frac{{\rm d}}{{\rm d} t} e^{X(t)} \right)e^{-X}&=&\frac{{\rm d}}{{\rm d} t}X(t) +\frac{1}{2}\left[X,\frac{{\rm d}}{{\rm d} t}X(t) \right]+\frac{1}{3!}\left[X, \,\left[X,\,\frac{{\rm d}}{{\rm d} t}X(t)\right]\right]+\dots\nonumber
\eea
The result is
\be
\delta \zeta=\frac{1}{3}\pd_i\xi^i +\xi^k\pd_k\zeta\;.
\label{zetatransfm}
\ee

Substituting $\delta\zeta$ back into~(\ref{identity1}), we can then solve for $\delta \gamma$. Following~\cite{Hinterbichler:2013dpa},
we do so perturbatively in powers of $\gamma$, expanding the tensor variation and the diffeormophism as follows:
\bea 
\nonumber
\delta \gamma_{ij}&=& \delta \gamma_{ij}^{(0)}+ \delta \gamma_{ij}^{(1)}+\ldots\\
\xi_i&=&\xi_i^{(0)}+\xi_i^{(1)}+\ldots
\label{tensexpand}
\eea
At zeroth-order in $\gamma$, the result is
\be
\delta \gamma^{(0)}_{ij}=\pd_i\xi_j^{(0)}+\pd_j\xi_i^{(0)}-\frac{2}{3}\pd^k \xi^{(0)}_k \delta_{ij}\;,
\label{delgam0}
\ee
which is manisfestly traceless. To ensure $\delta\gamma^{(0)}$ is also transverse, $\xi^{(0)}$ must satisfy
\be
\na^2 \xi^{(0)}_i+\frac{1}{3}\pd_i \pd^k \xi^{(0)}_k = 0 \;.
\label{xi0eqn}
\ee
As a check, note that if we assume that $\xi^{(0)}$ falls off at spatial infinity, then it is easy to see that the unique solution
is $\xi^{(0)} = 0$. This confirms that the gauge is completely fixed with respect to these boundary conditions. 

There are, however, non-trivial solutions of~(\ref{xi0eqn}) with $\xi^{(0)}$ diverging at infinity. In general, the solution is given by a power series\footnote{The constant term in the series,
representing an unbroken spatial translation, has been ignored. It is linearly realized and hence does not lead to
a soft-pion theorem.}:
\be 
\xi_i^{(0)} = \sum_{n=0}^\infty \frac{1}{(n+1)!}M_{i\ell_0 \ldots \ell_n}  x^{\ell_0}\cdots x^{\ell_n}\,,
\label{xiexpansion}
\ee
where the array $M_{i\ell_0\cdots \ell_n}$ is constant and symmetric in its last $n+1$ indices. To satisfy~(\ref{xi0eqn}), the array must obey the trace condition
\be
M_{i jj \ell_2\ldots \ell_n}  = - \frac{1}{3} M_{j i j \ell_2\ldots \ell_n}\qquad ({\rm for}~{\rm all}~n \geq 1) \,.
\label{Mcond}
\ee
Such transformations map field configurations that fall off at infinity into those that do not. However, certain linear combinations of them can be smoothly
extended to physical configurations with suitable fall-off behavior~\cite{Hinterbichler:2013dpa}, {\it i.e.}, these correspond to {\it adiabatic modes}~\cite{Weinberg:2003sw}. 
In particular, the spatial diffeomorphisms~(\ref{xiexpansion}) must be {\it time-independent}.\footnote{Strictly speaking, it must also be supplemented by a
time-dependent transformation, but this has no impact on the Ward identities at the end of the day. See~\cite{Hinterbichler:2013dpa} for a detailed discussion.} 

A further adiabatic restriction comes from the requirement that the non-linear shift $\delta\gamma_{ij}^{(0)}$ should remain transverse
when extended to a physical mode, {\it i.e.}, with smooth profile around $\vec{q} = 0$. To ensure that transversality is preserved at finite momentum,
$\hat{q}^i {\delta} \gamma_{i\ell_0}(\vec{q}) = 0$, the $M_{i\ell_0\cdots \ell_n}$ coefficients must become $\hat{q}$-dependent,
such that~\cite{Hinterbichler:2013dpa}
\be
\hat{q}^i  \left( M_{i\ell_0\ell_1 \ldots \ell_n}(\hat{q}) + M_{\ell_0 i\ell_1 \ldots \ell_n}(\hat{q}) - \frac{2}{3}\delta_{i\ell_0} M_{\ell\ell\ell_1 \ldots \ell_n}(\hat{q})\right) = 0\,.
\label{Mtrans}
\ee

Moving on to first-order in $\gamma$, the tensor variation is
\be
\delta \gamma^{(1)}_{ij}=\pd_i\xi_j^{(1)}+\pd_j\xi_i^{(1)}-\frac{2}{3}\pd^k \xi^{(1)}_k \delta_{ij}
+ \frac{1}{2} \left({\cal L}_{\xi^{(0)}}\gamma_{ij} + \xi^{(0)}_k\pd_k\gamma_{ij} - \pd_k\xi^{(0)}_i\gamma_{kj}- \pd_k\xi^{(0)}_j\gamma_{ki}\right) \;, 
\label{delta gamma1}
\ee
where ${\cal L}_{\xi^{(0)}}\gamma_{ij} \equiv\xi^{(0)}_k\pd_k\gamma_{ij}  +  \pd_i \xi^{(0)}_k\gamma_{kj} +\pd_j \xi^{(0)}_k\gamma_{ki}$ is the Lie derivative along $\xi^{(0)}$.
Imposing transversality of $\delta\gamma^{(1)}$ allows us to solve for $\xi^{(1)}$. Since $\xi^{(1)} \sim \gamma$, and $\gamma$ falls off at infinity, we can invert Laplancians
assuming fall-off boundary conditions. The solution is
\be
\xi^{(1)}_i=\frac{1}{2}\frac{\pd_j}{\na^2} \left(
\delta_{ik}-\frac{1}{4}\frac{\pd_i\pd_k}{\na^2}
\right)
\left(-{\cal L}_{\xi^{(0)}}\gamma_{kj}
+\pd_m \xi^{(0)}_k\gamma_{mj}
+\pd^\ell \xi^{(0)}_\ell\gamma_{kj} 
\right)\,.
\label{xi1}
\ee
This procedure can be straightforwardly extended to all orders in $\gamma$.

Combining the above results, the momentum-space field variations to first order $\gamma$ are given by
\bea
\nonumber
\delta \zeta(\vec{k}) &=& \sum_{n=0}^\infty \frac{(-i)^n}{3n!} M_{ii\ell_1 \ldots \ell_n}\frac{\partial^{n}}{\partial k_{\ell_1}\cdots \partial k_{\ell_n}}\left((2\pi)^3\delta^3(\vec{k})\right) \\
\nonumber
& -&  \sum_{n=0}^\infty\frac{(-i)^n}{n!} M_{i \ell_0 \ldots \ell_n} \Bigg( \delta^{i\ell_0} \frac{\partial^{n}}{\partial k_{\ell_1}\cdots \partial k_{\ell_n}} + \frac{k^{i}}{n+1}  \frac{\partial^{n+1}}{\partial k_{\ell_0} \cdots \partial k_{\ell_n}}\Bigg)\zeta(\vec{k}) \\
\nonumber
&+&  \sum_{n=0}^\infty\frac{(-i)^n}{n!} M_{\ell \ell_0\ldots \ell_n}\Upsilon^{\ell\ell_0ij}(\hat{k}) \frac{\partial^{n}}{\partial k_{\ell_1}\cdots \partial k_{\ell_n}}  \gamma_{ij}(\vec{k}) + \ldots \\
\nonumber
\delta \gamma_{ij}(\vec{k}) &=&  \sum_{n=0}^\infty\frac{(-i)^n}{n!} \left( M_{ij\ell_1 \ldots \ell_n} + M_{j i\ell_1 \ldots \ell_n} - \frac{2}{3}\delta_{ij} M_{\ell\ell\ell_1 \ldots \ell_n}\right) \frac{\partial^{n}}{\partial k_{\ell_1}\cdots \partial k_{\ell_n}}\left((2\pi)^3\delta^3(\vec{k})\right) \\
\nonumber
&-&   \sum_{n=0}^\infty\frac{(-i)^n}{n!} M_{\ell \ell_0\ldots \ell_n}\Bigg(\delta^{\ell\ell_0} \frac{\partial^{n}}{\partial k_{\ell_1}\cdots \partial k_{\ell_n}} + \frac{k^{\ell}}{n+1}\frac{\partial^{n+1}}{\partial k_{\ell_0} \cdots \partial k_{\ell_n}}\Bigg) \gamma_{ij}(\vec{k})\\
&+&   \sum_{n=0}^\infty\frac{(-i)^n}{n!} M_{ab \ell_1 \ldots \ell_n}  \Gamma^{ab}_{\;\;\;ij k\ell}(\hat{k}) \frac{\partial^{n}}{\partial k_{\ell_1}\cdots \partial k_{\ell_n}}  \gamma^{k\ell}(\vec{k}) + \ldots
\label{delnmom}
\eea
where the ellipses indicate higher-order terms in $\gamma$, and
\bea
\nonumber
\Upsilon_{abcd}(\hat{k})  &\equiv & \frac{1}{4}\delta_{ab}\hat{k}_c\hat{k}_d -\frac{1}{8}\delta_{ac}\hat{k}_b\hat{k}_d -\frac{1}{8}\delta_{ad}\hat{k}_b\hat{k}_c\,;\\
\Gamma_{a b i j k \ell}(\hat k)&\equiv & -\frac{1}{2} \left(\delta_{ij}+{\hat k}_i{\hat k}_j\right) \left(\delta_{ab}{\hat k}_k{\hat k}_\ell -\frac{1}{2}\delta_{a k}{\hat k}_\ell{\hat k}_b -\frac{1}{2}\delta_{a \ell}{\hat k}_k{\hat k}_b \right)
+\delta_{b(i}\delta_{j)(k}\delta_{\ell)a} -\delta_{a(i}\delta_{j)(k}\delta_{\ell)b} \nonumber \\
&-& \delta_{b(i}{\hat k}_{j)}\delta_{a (k } {\hat k}_{\ell)} +\delta_{a(i}{\hat k}_{j)}\delta_{b (k } {\hat k}_{\ell)} -\delta_{a (k }\delta_{\ell) (i}{\hat k}_{j)}{\hat k}_b -\delta_{b (k }\delta_{\ell) (i}{\hat k}_{j)}{\hat k}_a
+2\delta_{a b} {\hat k}_{(i}\delta_{j)(k}{\hat k}_{\ell)}\,.
\label{Gamdef}
\eea

\section{Derivation of the Consistency Conditions}
\label{derivereview}

In this Section, we briefly review the derivation of the consistency relations as Ward identities for the non-linearly realized symmetries discussed above. The full derivation can be
found in~\cite{Hinterbichler:2013dpa}, but we review it here both for completeness and by means of highlighting the key steps.

The starting point is the in-vacuum expectation value of the action of the conserved charge $Q$ on some operator ${\cal O}$:
\beq
\la \Omega | [Q,\mathcal{O}] | \Omega \ra=-i \la\Omega | \delta \mathcal{O}| \Omega \ra\,,
\label{wardstart}
\eeq
where ${\cal O}$ denotes an arbitrary equal-time product of $\zeta$'s and $\gamma$'s, and $|\Omega \ra$ is the in-vacuum of the interacting theory. 
The charge can be decomposed as
\be
Q = Q_0 + W\,,
\ee
where $Q_0$ generates the non-linear part of the transformations~(\ref{delnmom}), and $W$ is the remainder. By definition, $Q_0$ is a
symmetry of the free theory ({\it i.e.}, of the quadratic action). Specifically, for the field variations~(\ref{delnmom}), 
\be
Q_0 =  \lim_{\vec{q}\rightarrow 0} \sum_{n= 0}^\infty \frac{(-i)^n}{n!} M_{i\ell_0 \ldots \ell_n}(\hat{q}) \frac{\partial^{n}}{\partial q_{\ell_1}\cdots \partial q_{\ell_n}} \left( \frac{1}{3} \delta^{i\ell_0} \Pi_\zeta(\vec{q}) + 2\Pi_\gamma^{i\ell_0}(\vec{q}) \right)\,,
\label{Q0reg2}
\ee
where $\Pi_\zeta$ and $\Pi_\gamma^{ij}$ are the conjugate momenta.

By definition, the in-vacuum is related to the free vacuum by $\vert \Omega\rangle = \Omega(-\infty) \vert 0 \rangle$, 
where $\Omega(-\infty) \equiv U^\dagger (-\infty,0) U_0(-\infty,0)$, with $U$ and $U_0$ denoting respectively the full and free
time evolution operators. Similarly, 
\be
Q  \,  \Omega(-\infty)  =  \Omega(-\infty) Q_0\,. 
\label{Qident}
\ee
It follows that
\be
\label{QUU1}
Q |\Omega \rangle  =  Q \,\Omega(-\infty) \vert 0 \rangle = \Omega(-\infty)  Q_0 | 0 \rangle \, .
\ee
The action of the free charge $Q_0$ on the free vacuum $ | 0 \rangle$ can be computed by inserting a complete set of free-field 
eigenstates $|\zeta_0,\gamma_0\rangle$:
\begin{eqnarray}
\label{wavefunc0}
Q_0 |0 \rangle &=& \int D\zeta_0 D\gamma_0 \, \vert
\zeta_0, \gamma_0 \rangle \langle \zeta_0, \gamma_0\vert Q_0  \vert 0 \rangle
\nonumber \\
&=&  \lim_{\vec{q}\rightarrow 0} \sum_{n= 0}^\infty \frac{(-i)^n}{n!} M_{i\ell_0 \ldots \ell_n} \frac{\partial^{n}}{\partial q_{\ell_1}\cdots \partial q_{\ell_n}} \int D\zeta_0 D\gamma_0 \,
\vert \zeta_0, \gamma_0 \rangle \left\langle \zeta_0, \gamma_0 \left\vert \left( {1\over 3} \delta^{ij} \Pi_{\zeta_0}(\vec q) + 2 \Pi^{ij}_{\gamma_0}
  (\vec q) \right) \right\vert 0 \right\rangle
\nonumber \\
&=&  \lim_{\vec{q}\rightarrow 0} \sum_{n= 0}^\infty \frac{(-i)^{n+1}}{n!} M_{i\ell_0 \ldots \ell_n} \frac{\partial^{n}}{\partial q_{\ell_1}\cdots \partial q_{\ell_n}} \int D\zeta_0 D\gamma_0 \,
\vert \zeta_0, \gamma_0 \rangle
\left({1\over 3} \delta^{ij} {\delta \over \delta \zeta_0 (-\vec q)}
+2 {\delta \over \delta \gamma_0 {}_{ij} (-\vec q)}
\right) \langle \zeta_0, \gamma_0 \vert 0 \rangle \,. \nonumber \\
\end{eqnarray}
The free vacuum wavefunctional $\langle \zeta_0, \gamma_0 | 0 \rangle$ is a Gaussian
\be
\label{wavefunc1}
\langle \zeta_0, \gamma_0 | 0 \rangle \sim  \exp\left[ - \int \frac{{\rm d}^3k}{(2\pi)^3} \,\left(\frac{1}{4}
    \zeta_0(\vec{k})P_\zeta^{-1} (k)\zeta_0(-\vec{k}) +
    \frac{1}{8}\gamma_0 {}_{ij} (\vec{k}) P_\gamma^{-1} (k) \gamma_0 {}^{ij}(-\vec{k})\right) \right]\,,
\ee
up to an irrelevant phase which eventually drops out of the calculation. Substituting into~(\ref{wavefunc0}), we obtain
\be
\label{Qbar00}
Q_0 | 0 \rangle = -\lim_{\vec{q}\rightarrow 0} \sum_{n= 0}^\infty \frac{(-i)^{n+1}}{2n!} M_{i\ell_0 \ldots \ell_n} \frac{\partial^{n}}{\partial q_{\ell_1}\cdots \partial q_{\ell_n}} 
\left( {\delta^{ij} \over 3}  \frac{\zeta_0 (\vec q)}{P_{\zeta}(q)} +  \frac{\gamma_0^{\,ij} (\vec q)}{P_{\gamma}(q)} \right) \vert 0 \rangle \, .
\ee

At this point comes the assumption that the fields $\zeta$ and $\gamma$ are conserved operators as $\vec{q}\rightarrow 0$. In other words, we assume that\footnote{This statement is quite subtle, since
in~(\ref{wavefunc0}) we are instructed to differentiate $\zeta(\vec{q})$ and $\gamma(\vec{q})$ before sending $\vec{q}\rightarrow 0$. We refer the reader to~\cite{Hinterbichler:2013dpa} for a rigorous justification of this result.}
\be
\lim_{\vec{q}\rightarrow 0} \zeta(\vec{q}) \,  \Omega(-\infty) = \lim_{\vec{q}\rightarrow 0}  \Omega(-\infty) \zeta_0 (\vec{q})  \,;\qquad \lim_{\vec{q}\rightarrow 0} \gamma^{ij}(\vec{q}) \,  \Omega(-\infty) = \lim_{\vec{q}\rightarrow 0}  \Omega(-\infty) \gamma^{ij}_0 (\vec{q}) \,.
\ee
Combined with~(\ref{Qident}), this allows us to rewrite~(\ref{Qbar00}) in terms of the operators of the {\it full} interacting theory:
\be
Q | \Omega \rangle = -\lim_{\vec{q}\rightarrow 0} \sum_{n= 0}^\infty \frac{(-i)^{n+1}}{2n!} M_{i\ell_0 \ldots \ell_n} (\hat{q})\frac{\partial^{n}}{\partial q_{\ell_1}\cdots \partial q_{\ell_n}} 
\left( {\delta^{ij} \over 3}  \frac{\zeta (\vec q)}{P_{\zeta}(q)} +  \frac{\gamma^{\,ij} (\vec q)}{P_{\gamma}(q)} \right) \vert \Omega \rangle \, .
\ee
The left-hand side of the Ward identity~(\ref{wardstart}) therefore reduces to
\be
\langle \Omega \vert [Q, {\cal O}]  \vert\Omega \rangle = \lim_{\vec{q}\rightarrow 0} \sum_{n= 0}^\infty \frac{(-i)^{n+1}}{n!} M_{i\ell_0 \ldots \ell_n}(\hat{q}) \frac{\partial^{n}}{\partial q_{\ell_1}\cdots \partial q_{\ell_n}} \Bigg(\frac{ \la \gamma^{i\ell_0}(\vec{q}){\cal O}\ra }{P_\gamma(q)}+ \frac{\delta^{i\ell_0}}{3} \frac{\la \zeta(\vec{q}) {\cal O} \ra}{P_\zeta(q)}  \Bigg)\,. 
\label{lhsmostgeneral}
\ee

Meanwhile, the right-hand side, $-i \la\Omega | \delta \mathcal{O}| \Omega \ra$, is just given by the field variations~(\ref{delnmom}) of the various $\zeta$'s and $\gamma$'s implicit in ${\cal O}$. 
As argued in~\cite{Hinterbichler:2013dpa}, the non-linear parts of $\delta\zeta$ and $\delta\gamma$, given by derivatives of delta functions in~(\ref{delnmom}) contribute to disconnected diagrams
in the Ward identity. Focusing on connected correlators, denoted by $\la\ldots\ra_c$, the Ward identities take the form
\be
\begin{split}\label{rhsfinal}
 \lim_{\vec{q}\rightarrow 0}& M_{i\ell_0 \ldots \ell_n}(\hat{q}) \frac{\partial^{n}}{\partial q_{\ell_1}\cdots \partial q_{\ell_n}} \Bigg(\frac{1}{P_\gamma(q)} \la \gamma^{i\ell_0}(\vec{q}){\cal O}(\vec{k}_1,\ldots,\vec{k}_N) \ra_c + \frac{\delta^{i\ell_0}}{3P_\zeta(q)} \la  \zeta(\vec{q}) {\cal O}(\vec{k}_1,\ldots,\vec{k}_N) \ra_c  \Bigg) \\
 = &-  \sum_{n=0}^\infty \frac{(-i)^{n}}{n!} M_{i\ell_0 \ldots \ell_n}(\hat{q}) \Bigg\{ \sum_{a=1}^N\Bigg( \delta^{i\ell_0} \frac{\partial^{n}}{\partial k_{\ell_1}^a\cdots \partial k_{\ell_n}^a} + \frac{k^{i}_a}{n+1}   \frac{\partial^{n+1}}{\partial k_{\ell_0}^a \cdots \partial k_{\ell_n}^a}\Bigg)\la {\cal O}(\vec{k}_1,\ldots,\vec{k}_N) \ra_c \\
& -\sum_{a=1}^M \Upsilon^{i\ell_0i_aj_a}(\hat{k}_a)  \frac{\partial^{n}}{\partial k_{\ell_1}^a\cdots \partial k_{\ell_n}^a} \la {\cal O}^\zeta(\vec{k}_1,\ldots,\vec{k}_{a-1},\vec{k}_{a+1},\ldots \vec{k}_M)\gamma_{i_aj_a}(\vec{k}_a) {\cal O}^\gamma (\vec{k}_{M+1},\ldots,\vec{k}_N)\ra_c \\
& - \sum_{b=M+1}^N  \Gamma^{i\ell_0\;\;\;\;\;k_b\ell_b}_{\;\;\;\;i_bj_b}(\hat{k}_b)\frac{\partial^{n}}{\partial k_{\ell_1}^b\cdots \partial k_{\ell_n}^b} \la {\cal O}^\zeta(\vec{k}_1,\ldots,\vec{k}_M) {\cal O}^\gamma_{i_{M+1} j_{M+1},\ldots,k_b\ell_b,\ldots i_Nj_N}(\vec{k}_{M+1},\ldots,\vec{k}_N) \ra_c \Bigg\}\\
& + \ldots
\end{split}
\ee
where, as before, the ellipses indicate higher-order terms in $\gamma$. Here, ${\cal O}^\zeta$ and ${\cal O}^\gamma$ respectively denote products of $\zeta$'s and $\gamma$'s.

Finally, it is convenient to express the identity in terms of on-shell correlation functions, obtained by removing the momentum-conserving delta functions:
\be
\la {\cal O}(\vec{k}_1, \ldots ,\vec{k}_N)\ra  = (2\pi)^3\delta^3(\vec{K}_t) \la {\cal O}(\vec{q}, \vec{k}_1, \ldots ,\vec{k}_N)\ra' \,,
\label{primecor}
\ee
where $\vec{K}_t \equiv \vec{k}_1 + \ldots + \vec{k}_N$ is the total momentum. Removing the delta functions involves some technical subtleties, which are explained in detail in Appendix~B.
Our convention for going on-shell is to express $\vec{k}_N$ in terms of the $\vec{k}$'s, that is, $\vec k_N = -\vec k_1 -\dots -\vec k_{N-1}$. We then distinguish two cases for the Ward identities in terms of primed correlators:

\begin{itemize}

\item {Hard modes containing at least one $\gamma$ field:} In this case, the primed identities are given by:
\bea
\nonumber
 & & \lim_{\vec{q}\rightarrow 0} M_{i\ell_0 \ldots \ell_n}(\hat{q}) \frac{\partial^{n}}{\partial q_{\ell_1}\cdots \partial q_{\ell_n}} \Bigg(\frac{1}{P_\gamma(q)} \la \gamma^{i\ell_0}(\vec{q}){\cal O}(\vec{k}_1,\ldots,\vec{k}_N) \ra_c' + \frac{\delta^{i\ell_0}}{3P_\zeta(q)} \la  \zeta(\vec{q}) {\cal O}(\vec{k}_1,\ldots,\vec{k}_N) \ra_c'  \Bigg) \\
\nonumber
& & ~~ =  - M_{i\ell_0 \ldots \ell_n}(\hat{q}) \Bigg\{ \sum_{a=1}^{N-1} \Bigg( \delta^{i\ell_0} \frac{\partial^{n}}{\partial k_{\ell_1}^a\cdots \partial k_{\ell_n}^a} - \frac{\delta_{n0}}{N-1}\delta^{i\ell_0}
+ \frac{k^{i}_a}{n+1}  \frac{\partial^{n+1}}{\partial k_{\ell_0}^a \cdots \partial k_{\ell_n}^a}\Bigg) \la  {\cal O}(\vec{k}_1,\ldots,\vec{k}_N) \ra_c'  \\
\nonumber
& & \;\;\;\;~~-\sum_{a=1}^M \Upsilon^{i\ell_0i_aj_a}(\hat{k}_a) \Big\vert_{\rm OS}\; \frac{\partial^{n}}{\partial k_{\ell_1}^a\cdots \partial k_{\ell_n}^a} \la {\cal O}^\zeta(\vec{k}_1,\ldots,\vec{k}_{a-1},\vec{k}_{a+1},\ldots \vec{k}_M)\gamma_{i_aj_a}(\vec{k}_a) {\cal O}^\gamma (\vec{k}_{M+1},\ldots,\vec{k}_N)\ra_c' \\
\nonumber
& & \;\;\;\; ~~ -\sum_{b=M+1}^{N-1}  \Gamma^{i\ell_0\;\;\;\;\;k_b\ell_b}_{\;\;\;\;i_bj_b}(\hat{k}_b )\Big\vert_{\rm OS}\;\frac{\partial^{n}}{\partial k_{\ell_1}^b\cdots \partial k_{\ell_n}^b}\la {\cal O}^\zeta(\vec{k}_1,\ldots,\vec{k}_M) {\cal O}^\gamma_{i_{M+1} j_{M+1},\ldots,k_b\ell_b,\ldots i_Nj_N}(\vec{k}_{M+1},\ldots,\vec{k}_N) \ra_c' \\ 
\nonumber
& & \;\;\;\; ~~ -(-1)^n
\frac{\partial^{n}}{\partial k_{\ell_1}^N\cdots \partial k_{\ell_n}^N}\Gamma^{i\ell_0\;\;\;\;\;~~k_N\ell_N}_{\;\;\;\;i_N j_N}(\hat{k}_N ) \Bigg\vert_{\rm OS}\nonumber \\
& &\;\;\;\;~~\times \la {\cal O}^\zeta(\vec{k}_1,\ldots,\vec{k}_M) {\cal O}^\gamma_{i_{M+1} j_{M+1},\ldots, i_{N-1}j_{N-1}, k_N\ell_N}(\vec{k}_{M+1},\ldots,\vec{k}_N) \ra_c' 
\Bigg\} + \ldots
\label{wardalmosttheren>2}
\eea
where ``OS" stands for ``on-shell".

\item{Hard modes consisting of scalars only:} In this case, the primed identities are given by:
\bea
\nonumber
 & & \lim_{\vec{q}\rightarrow 0} M_{i\ell_0 \ldots \ell_n}(\hat{q}) \frac{\partial^{n}}{\partial q_{\ell_1}\cdots \partial q_{\ell_n}} \Bigg(\frac{1}{P_\gamma(q)} \la \gamma^{i\ell_0}(\vec{q}){\cal O}(\vec{k}_1,\ldots,\vec{k}_N) \ra_c' + \frac{\delta^{i\ell_0}}{3P_\zeta(q)} \la  \zeta(\vec{q}) {\cal O}(\vec{k}_1,\ldots,\vec{k}_N) \ra_c'  \Bigg) \\
\nonumber
& & ~~ =  - M_{i\ell_0 \ldots \ell_n}(\hat{q}) \Bigg\{ \sum_{a=1}^{N-1} \Bigg( \delta^{i\ell_0} \frac{\partial^{n}}{\partial k_{\ell_1}^a\cdots \partial k_{\ell_n}^a} - \frac{\delta_{n0}}{N-1}\delta^{i\ell_0}
+ \frac{k^{i}_a}{n+1}  \frac{\partial^{n+1}}{\partial k_{\ell_0}^a \cdots \partial k_{\ell_n}^a}\Bigg) \la  {\cal O}(\vec{k}_1,\ldots,\vec{k}_N) \ra_c'  \\
\nonumber
& & \;\;\;\;~~-\sum_{a=1}^{N-1} \Upsilon^{i\ell_0i_aj_a}(\hat{k}_a) \Big\vert_{\rm OS}\; \frac{\partial^{n}}{\partial k_{\ell_1}^a\cdots \partial k_{\ell_n}^a} \la {\cal O}^\zeta(\vec{k}_1,\ldots,\vec{k}_{a-1},\vec{k}_{a+1},\ldots \vec{k}_N)\gamma_{i_aj_a}(\vec{k}_a) \ra_c'\nonumber\\ 
& & \;\;\;\;~~-(-1)^n 
\frac{\partial^{n}}{\partial k_{\ell_1}^N\cdots \partial k_{\ell_n}^N}\Upsilon^{i\ell_0i_Nj_N}(\hat{k}_N) \Bigg\vert_{\rm OS}\;  \la {\cal O}^\zeta(\vec{k}_1,\ldots,\vec{k}_{N-1})\gamma_{i_N j_N}(\vec{k}_N)\ra_c'
\Bigg\} + \ldots
\eea

\end{itemize}

Note that the $\Upsilon$ term replaces every $\zeta$ in ${\cal O}$ with a $\gamma$ insertion, whereas the $\Gamma$ term replaces every $\gamma$ in ${\cal O}$ with another $\gamma$, with suitably contracted indices. The $\partial \Gamma/\partial k$ terms were missed in~\cite{Hinterbichler:2013dpa} and arise from a careful removal of the momentum-conserving delta functions. See Appendix~B for a detailed discussion of this point.

\section{Three Point Functions}
\label{3ptcalc}

In this work, we are interested in single-clock models of inflation with arbitrary scalar sound speed $c_s\neq 1$. These models can be described at once within the framework of the effective theory of inflation~\cite{Cheung:2007sv}. In this approach, the theories at hand are conveniently expressed in terms of $\pi$, the St\"uckelberg field for broken time translational invariance.
Ignoring terms involving the extrinsic curvature perturbation $\delta K^\mu_{~\nu}$ for simplicity, the effective action is given by
\begin{align}
\label{EFT of Inflation}
S&= \int {\rm d}^4x \sqrt{-g} \bigg [ \frac{M_{\rm Pl}^2}{2}R - M_{\rm Pl}^2\left(3H(t+\pi)^2+\dot{H}(t+\pi)\right)
\nonumber\\
&\quad~~~~~~~~~ -\mpl^2 \dot{H}(t+\pi) Q +\frac{1}{2}M(t+\pi)^4 Q^2 +\frac{1}{6}c_3(t+\pi)M(t+\pi)^4Q^3 + \ldots \bigg]\,,
\end{align}
where $N$ and $N^i$ denote the lapse function and shift vector of the ADM decomposition, and
\be
Q\equiv \frac{1}{N^2}\left(1+\dot{\pi}-N^i\partial _i \pi\right)^2-h^{ij}\partial_i\pi\partial_j\pi-1\,. 
\ee
Notice that the action \eqref{EFT of Inflation} is invariant under the spatial diffeomorphsms as well as the ``diagonal'' time diffeomorphism: $t\to t+\xi^0(x)\;,\;\pi\to \pi-\xi^0(x)$.
Although we have dropped extrinsic curvature contributions, the above effective theory is sufficiently general to perform non-trivial checks of the consistency relations. 

Eventually we are interested in 3-point correlators in $\zeta$-gauge specified by~\eqref{zeta gij}. In order to check the Ward identities \eqref{wardalmosttheren>2}, we need to compute the correlators of $\zeta$ field itself as well as those involving both $\zeta$ and $\gamma$ fields, at leading order in slow roll. It turns out that, for this purpose, the computation can be most conveniently performed in spatially-flat gauge, defined by
\be
\phi(x)=\bar{\phi}\big(t+\pi(x)\big)\;,\quad h_{ij}=a^2(t) \left(e^{\tilde \gamma}\right)_{ij}\;,\quad \pd_j{\tilde \gamma}_{ij}=\tilde{\gamma}_{ii}=0\;.
\ee 
Then once we know the correlators of $\pi$ and $\tilde{\gamma}$, we can translate them into those of $\zeta$ and $\gamma$ via
\be
\zeta \simeq-H\pi\;; \quad {\tilde \gamma}_{ij}\simeq \gamma_{ij}\,,
\ee
which are valid to leading order in slow roll.

Therefore all we need for our computation is the action up to the cubic order in perturbations. This requires solving the constraint equations for the lapse and shift to
linear order~\cite{Maldacena:2002vr,Cheung:2007sv}
\beq
\nonumber
\delta N &\equiv& N - 1= \epsilon H \pi\,;\\
\partial^i N_i&=&- \frac{\epsilon }{c_s^2}\frac{\partial}{\partial t}(H \pi)\,,
\eeq
where $\epsilon \equiv -\dot{H}/H^2$, and the sound speed $c_s$ is related to the coefficients in~\eqref{EFT of Inflation} by
\beq
c_s^{-2}=1-\frac{2M^4(t)}{\dot{H}\mpl ^2}\,.
\eeq

Substituting for $\delta N$ and $N^i$ into~(\ref{EFT of Inflation}), we obtain the following quadratic Lagrangian density
\beq
\mathcal{L}^{(2)}&=&\frac{\mpl^2}{8} a^3 \left( \dot{\gamma}_{ij} \dot{\gamma}_{ij} -\frac{1}{a^2} \pd _i \gamma_{jk} \pd_{i}\gamma_{jk} \right) \nonumber \\
&+&\frac{\mpl^2a^3H^2\epsilon}{c_s^2}\left( \dot{\pi}^2-\frac{c_s^2}{a^2}\pd_i\pi\pd_i\pi+H^2\epsilon \pi^2 \Big( 3-2(s+\epsilon-\eta) \Big) \right)\,,
\label{L2}
\eeq
where, as usual, other slow roll parameters are defined by $\eta \equiv H^{-1} {\rm d}\ln\epsilon/{\rm d}t$ and $s\equiv H^{-1} {\rm d}\ln c_s/{\rm d}t$. We also assume that they are of the same order as $\epsilon$: $\eta\sim s \sim {\cal O}(\epsilon)$.\footnote{In fact $\eta, s$ will not appear in the three point functions at leading order in slow roll.
}
The 2-point functions for $\zeta$ and $\gamma$ can be readily obtained from~\eqref{L2}. To leading order in the slow roll approximation,
they are given by
\be
\la \zeta_{\vec{k}} \zeta_{-\vec{k}}  \ra ' = \frac{H^2}{4 \epsilon c_s \mpl^2k^3}\,;\qquad \la \gamma^{ij}_{\vec{k}} \gamma^{kl}_{-\vec{k}}  \ra '=\frac{H^2}{\mpl^2 k^3}  \Pi^{ijk\ell}(\hat{k}) \,,
\ee
where
\be
\Pi_{ijk\ell}(\hat k)\equiv \sum_{s=\pm}e^s_{ij}(\hat k)e^s_{k\ell}(\hat k)
= P_{ik}P_{j\ell}+P_{i\ell}P_{jk}-P_{ij}P_{k\ell}\,;\qquad P_{ij}(\hat k)\equiv\delta_{ij}-\hat{k}_i\hat{k}_j\,.
\label{Pidef}
\ee

The cubic Lagrangian density for the perturbations is given by
\beq
\mathcal{L}^{(3)}&=&\mathcal{L}_{\pi\pi\pi}+\mathcal{L}_{\gamma\pi\pi}+\mathcal{L}_{\gamma\gamma\pi}+\mathcal{L}_{\gamma\gamma\gamma}\,,
\eeq
with
\beq
\nonumber
\mathcal{L}_{\pi\pi\pi}&=&a^3\frac{\mpl^2H^2\epsilon}{c_s^2}\left( C_{\dot{\pi}^3} \dot{\pi}^3+C_{\dot{\pi}(\pd\pi)^2}\frac{1}{a^2}\dot{\pi}(\pd\pi)^2+C_{\pi\dot{\pi}^2} \pi\dot{\pi}^2+C_{\pi(\pd\pi)^2}\frac{1}{a^2}\pi(\pd\pi)^2+C_{\rm NL} \dot{\pi} \pd_i \pi \pd^i\frac{1}{\vec{\nabla}^2}\dot{\pi}\right)\,; \\
\nonumber
\mathcal{L}_{\gamma\pi\pi}&=&\frac{\mpl^2}{2}a^3\left( \pd^j N^{i} \pd_m \gamma_{ij} N^{m}+\frac{1}{2}\delta N \dot{\gamma_{ij}}(\pd_i N^j+\pd_j N^i)\right)+\mpl^2 H^2 a\epsilon \gamma_{ij} \pd_i\pi\pd_j\pi\,;\\
\nonumber
\mathcal{L}_{\gamma\gamma\pi}&=&\frac{\mpl^2}{2}a^3\left( -\frac{1}{2}\dot{\gamma_{ij}}\pd_m \gamma_{ij}N^m-\frac{1}{4}\delta N\dot{\gamma_{ij}} \dot{\gamma_{ij}} -\frac{\delta N}{4 a^2} \pd_i\gamma_{jk} \pd_i\gamma_{jk} \right)\,;\\
\mathcal{L}_{\gamma\gamma\gamma}&=&\frac{\mpl^2}{2}a^3\left( \frac{1}{4a^2}\gamma_{im}\pd_i\gamma_{jk}\pd_m\gamma_{jk}-\frac{1}{2a^2}\gamma_{im}\pd_i\gamma_{kj}\pd_j\gamma_{km} \right)\,,
\eeq
The coefficients in ${\cal L}_{\pi\pi\pi}$ are related to the parameters of the effective theory, up to next-to-leading order in slow roll, by
\beq
\nonumber
C_{\dot{\pi}^3}&=&(1-c_s^2)\left(1+\frac{2}{3}c_3\right)\,;\\
\nonumber
C_{\dot{\pi}(\pd\pi)^2}&=&-1+c_s^2\,;\\
\nonumber
C_{\pi\dot{\pi}^2}&=&H\left(-6\epsilon+\eta-2s+3\epsilon c_s^2-2\epsilon c_3(1-c_s^2)\right)\,;\\
\nonumber
C_{\pi(\pd\pi)^2}&=&H\left(\epsilon-\eta c_s^2\right)\,; \\
C_{\rm NL}&=&\frac{2\epsilon H}{c_s^2}\,.
\eeq

Using this cubic action, we can compute various tree-level 3-point functions in the in-in formalism as usual. The results are:

\begin{itemize} 

\item {\bf Three scalars}:

\begin{align}
\langle \zeta_{\vec k_1} \zeta_{\vec k_2} \zeta_{\vec k_3} \rangle' 
=\frac{H^4(1-c_s^2)}{32 \mpl^4c_s^4 \ep^2}
\frac{12\left(1 -c_s^2 \left(1+\sfrac{2}{3}c_3\right)\right)  {\cal K}_2^6- 4 K {\cal K}_1^2 {\cal K}_2^3- 4K^2 {\cal K}_1^4+ 11 K^3 {\cal K}_2^3-3 K^4 {\cal K}_1^2+ K^6}{k_1^3 k_2^3 k_3^3 K^3} \,,
\label{3scalars}
\end{align}
where 
\bea
\nonumber
K &\equiv & k_1+k_2+k_3\;;\\ 
\nonumber
{\cal K}_1&=&\sqrt{k_1 k_2+k_1 k_3+k_2 k_3}\;;\\
{\cal K}_2&=&(k_1 k_2 k_3)^{1/3}\;.
\eea

\item {\bf Two scalars and one tensor}:

\begin{align}
\langle \gamma_{\vec p}^{ij} \zeta_{\vec k_1} \zeta_{\vec k_2} \rangle' =-\frac{H^4}{4\mpl^4\ep c_s^2} 
\frac{U(c_s k_1,c_s k_2,p)}{k_1^3 k_2^3 p^3} \Pi^{ij}_{\;~mn}(\hat{p})k_1^m k_2^n\;,
\label{1tensor2scalars}
\end{align}
where $\Pi_{ijmn}$ was defined in~(\ref{Pidef}), and 
\be
U(k_1, k_2, p)\equiv \frac{k_1^3+k_2^3+k_3^3+2k_1 k_2 k_3+\left(2k_1^2 k_2+ 5 \text{ perms}\right)}{K^2} \,
\ee

\item {\bf One scalar and two tensors}:

\be
\langle \gamma_{\vec p_1}^{ij} \gamma_{\vec p_2}^{k\ell} \zeta_{\vec k} \rangle' =\frac{H^4}{8\mpl^4 c_s } 
\frac{
\Pi^{ij}_{\;~mn}(\hat{p}_1)\Pi^{k\ell}_{\;~mn}(\hat{p}_2)
}
{k^3 p_1^3 p_2^3}
\left(
\frac{4 p_1^2 p_2^2}{c_s k +p_1+p_2}
-\frac{c_s k (k^2-p_1^2-p_2^2)}{2}
\right)
\label{1scalar2tensors}
\ee

\item {\bf Three tensors}:

The $\gamma\gamma\gamma$ correlator is of course identical to that in $c_s=1$ inflation model~\cite{Maldacena:2011nz}:
\be
\la\gamma^{ij}_{\vec p_1}\gamma^{k\ell}_{\vec p_2}\gamma^{m n}_{\vec p_3}\ra' =\frac{H^4}{2\mpl^4}
\frac{U(p_1,p_2,p_3)}{p_1^3 \,p_2^3 \,p_3^3}\Pi^{i j}_{\;~a a'}(\hat p_1)\Pi^{k \ell}_{ \;~b b'}(\hat p_2)\Pi_{m n}^{\;~c c'}(\hat p_3)t_{abc}t_{a' b' c'}\;,
\label{3tensors}
\ee
where 
\be
t_{a b c}=k_2^a \delta_{bc}+k_3^b\delta_{ac}+k_1^c\delta_{ab}\;.
\ee

\end{itemize}

The above 3-point functions, given by~\eqref{3scalars},~\eqref{1tensor2scalars},~\eqref{1scalar2tensors} and~\eqref{3tensors}, all agree with~\cite{Maldacena:2002vr} for $c_s =1$,
with the exception of a small typo\footnote{In Eq.~(4.13) of~\cite{Maldacena:2002vr}, given by 
\be
\langle \zeta_{\vec{k}_1} \gamma_{\vec{k}_2}^{s_2} \gamma_{\vec{k}_3}^{s_3} \rangle \sim \left(-\frac{1}{4}k_1^3 + \frac{1}{2} k_1\left(k_2^2+ k_3^2\right) +4\frac{k_2^2k_3^2}{K}\right) \,,
\ee
the $-1/4$ coefficient of the $k_1^3$ term should be $-1/2$. This is crucial checking the consistency relations with one scalar and one tensor insertion.}
It is important to stress, however, that the $c_s \neq 1$ results given here cannot simply be inferred by rescaling the $k$'s in the $c_s = 1$ correlators,
since the cubic Lagrangian involves extra non-trivial $c_s$ dependence as well. Indeed, we can see explicitly that the $\la\gamma \gamma \zeta \ra$ correlator \eqref{1scalar2tensors} can not be written as a function of $c_s k$, $p_1$ and $p_2$. (The illusion that $\la\gamma \zeta \zeta \ra$ correlator \eqref{1tensor2scalars} can is only an artifact of leading order in the slow roll approximation.)

\section{Consistency Relations with Two Scalars}
\label{zetazeta}

The simplest case is where the two hard modes (with momenta $\vec{k}_1$ and $\vec{k}_2$) are both scalars, {\it i.e.},
$\mathcal{O}(\vec{k}_1,\vec{k}_2)=\zeta_{\vec{k_1}}\zeta_{\vec{k_2}}$. In this case, the identity~(\ref{wardalmosttheren>2})
reduces to
\bea
\nonumber
 & & \lim_{\vec{q}\rightarrow 0} M_{i\ell_0 \ldots \ell_n}(\hat{q}) \frac{\partial^{n}}{\partial q_{\ell_1}\cdots \partial q_{\ell_n}} \Bigg(\frac{\la \gamma^{i\ell_0}_{\vec{q}} \zeta_{\vec{k_1}}\zeta_{\vec{k_2}} \ra'}{P_\gamma(q)}  + \frac{\delta^{i\ell_0}}{3} \frac{\la  \zeta_{\vec{q}} \zeta_{\vec{k_1}}\zeta_{\vec{k_2}} \ra'}{P_\zeta(q)} \Bigg) \\
& & ~~ =  - M_{i\ell_0 \ldots \ell_n}(\hat{q})\Bigg( \delta^{i\ell_0} \frac{\partial^{n}}{\partial k^{\ell_1}_1\cdots \partial k^{\ell_n}_1} + \frac{k^{i}_1}{n+1}  \frac{\partial^{n+1}}{\partial k^{\ell_0}_1 \cdots \partial k^{\ell_n}_1}\Bigg) P_\zeta(k_1)\;,\quad \text{for all }n\ge0\;.
\label{wardzetazeta}
\eea
Note that the $\Upsilon$ contribution in this case is proportional to $\langle \zeta\gamma\rangle$ and hence vanishes identically. The $\Gamma$ contribution is absent since the hard modes are scalars only.

In what follows we will check this identity explicitly for $0\leq n \leq 3$. For this purpose, we will need the 3-point functions~(\ref{3scalars}) and~(\ref{1tensor2scalars}) expanded to cubic order in the soft momentum. Note that we must first impose the on-shell condition before Taylor expanding. For instance, letting $\vec{k_1}\rightarrow \vec{q}$ in~(\ref{3scalars}), we first set $\vec{k}_3 = -\vec{k}_2 - \vec{q}$ and then expand in powers of $q$. The result is
\bea
\nonumber
\frac{\langle \zeta_{\vec{q}} \zeta_{\vec{k}_1} \zeta_{\vec{k}_2}\rangle}{P_\zeta(q)}  &=&  \frac{H^2(c_s^2-1)}{16M_{\rm Pl}^2\epsilon c_s^3k^3}\Bigg\{\frac{q^2}{k^2}\left( 8+ \left(3 + 2c_3 \right)c_s^2 - 5 \left(\hat{q}\cdot\hat{k}\right)^2 \right)\\
\nonumber
&-& \frac{q^3}{2k^3}\left(3\left(3 + (3 +2c_3)c_s^2\right) + 5 \hat{q}\cdot\hat{k} \left(10 + (3 +2c_3)c_s^2\right) - 35 \left(\hat{q}\cdot\hat{k}\right)^3\right)+ \ldots \Bigg\}  \\
\nonumber
\frac{\langle \gamma^{ij}_{\vec{q}} \zeta_{\vec{k}_1} \zeta_{\vec{k}_2}\rangle}{P_\gamma(q)} &=& \frac{H^2}{8M_{\rm Pl}^2 \epsilon c_s k^3} \Pi^{ij}_{\;~mn}(\hat{q}) \hat{k}^m\hat{k}^n\Bigg\{ 3 - \frac{15q}{2k} \hat{q}\cdot\hat{k}  + \frac{5q^2}{4k^2c_s^2}\left(1-3c_s^2  + 14c_s^2 \left(\hat{q}\cdot\hat{k}\right)^2\right) \\
&-& \frac{q^3}{8k^3c_s^3}\left(6 + 35c_s \left(1-4c_s^2\right)\hat{q}\cdot\hat{k} + 315 c_s^3 \left(\hat{q}\cdot\hat{k}\right)^3\right) + \ldots \Bigg\}\,,
\label{expand3pt}
\eea
where, for simplicity, we have denoted the hard momentum by $\vec{k}$. Since $\langle \zeta\zeta\zeta\rangle/P_\zeta$ starts at order $q^2$, it does not contribute to the $n=0$ and $n=1$ relations, as mentioned in
the Introduction. The $n=0$ and $n=1$ identities therefore boil down to the anisotropic and linear-gradient consistency relations~\cite{Creminelli:2012ed}.

Notice that the 3-point correlation functions involving a soft tensor mode are proportional to $\Pi_{ijk\ell}(\hat{q})$. This could, in principle, lead to the appearance of the derivatives of $\Pi_{ijk\ell}(\hat{q})$ on the left-hand side of \eqref{wardalmosttheren>2}. However, it was shown in \cite{Berezhiani:2013ewa} that 
\beq
M_{i\ell_0\ldots\ell_n}(\hat{q})\frac{\pd^n \Pi^{i\ell_0}_{\;~k\ell}(\hat q)}{\pd q_{\ell_1}\ldots \pd q_{\ell_n}}=0\,, \qquad  \text{for all }n\ge0\;,
\label{Mderproject}
\eeq
which follows solely from the properties of $M_{i\ell_0\ldots\ell_n}$. For this reason, we did not expand $\Pi_{ijk\ell}(\hat{q})$ in the Taylor series~(\ref{expand3pt}). 

\begin{itemize}

\item {\it Anisotropic scaling consistency relation} ($n=0$):  The identity in this case reduces to
\begin{align}
\lim_{\vec{q}\to 0}~M_{ij}(\hat{q})
	\frac{\la\gamma^{ij}_{\vec q}\zeta_{\vec k_1}\zeta_{\vec k_2}\ra'}{P_\gamma(q)}
= - M_{ij}(\hat{q}) k^i \frac{\pd}{\pd k^j} P_\zeta(k) \,,
\label{CR n=0}
\end{align}
where $M_{ij}$ can be assumed symmetric and traceless, without loss of generality. Thus it can be expressed as the linear combination of the polarization tensors, $M_{ij}(\hat q)=a e_{ij}^+(\hat q)+b e_{ij}^-(\hat q)$, and therefore
\be
M_{ij}(\hat q)\Pi^{ij}_{\;~mn}(\hat q)=2 M_{mn}(\hat q)\,.
\ee
Then the left-hand side of~\eqref{CR n=0} gives
\bea
\nonumber
\text{L.H.S.}&=&\lim_{q\to 0} M_{ij}(\hat{q})
	\frac{\la\gamma^{ij}_{\vec q}\zeta_{\vec k_1}\zeta_{\vec k_2}\ra'}{P_\gamma(q)} \\
&=& \frac{3H^2}{4\mpl^2 \epsilon c_s k^3 } M_{mn}(\hat q) \hat{k}^m \hat{k}^n\;.
\eea
This agrees with the right-hand side:
\bea
\nonumber
\text{R.H.S.}&=& - M_{ij}(\hat{q}) k^i \frac{\pd}{\pd k^j} P_\zeta(k) \\
&=&\frac{3H^2}{4\mpl^2 \epsilon c_s k^3 } M_{mn}(\hat q) \hat{k}^m \hat{k}^n\;.
\eea

\item {\it Linear-gradient consistency relation} ($n=1$): The identity at this order reduces to
\be
\lim_{\vec{q}\to 0} M_{ijm_1}(\hat{q}) \frac{\pd}{\pd q_{m_1}}\frac{\la\gamma^{ij}_{\vec q}\zeta_{\vec k_1}\zeta_{\vec k_2}\ra'}{P_\gamma(q)} 
\quad= -
	M_{ijm_1}(\hat{q})
	\left(
	\delta^{ij}\frac{\pd}{\pd k^{m_1}}
	+\frac{k^i}{2}\frac{\pd^2}{\pd k^j\pd k^{m_1}}
	\right)P_\zeta(k)\;.
\label{CR n=1}
\ee
When taking the derivative on the left-hand side, note that the term proportional to $\partial \Pi^{ij}_{\;~mn}/\partial q$ is projected out by $M_{ijm_1}(\hat{q})$, as mentioned in~(\ref{Mderproject}).
The result is
\bea
\nonumber
\text{L.H.S.}&=& \lim_{\vec{q}\to 0} M_{ijm_1}(\hat{q}) \frac{\pd}{\pd q_{m_1}}\frac{\la\gamma^{ij}_{\vec q}\zeta_{\vec k_1}\zeta_{\vec k_2}\ra'}{P_\gamma(q)} \\
&=& \frac{5H^2}{8\mpl^2 c_s \epsilon k^4}
	\left(
		 M_{jjm_1}(\hat q){\hat k}^{m_1}
		-3M_{m n m_1}(\hat q){\hat k}^{m}{\hat k}^{n}{\hat k}^{m_1}
	\right)\;.
\eea
This agrees with the right-hand side:
\bea
\nonumber
\text{R.H.S.}&=& - M_{ijm_1}(\hat{q}) \left( \delta^{ij}\frac{\pd}{\pd k^{m_1}}
	+\frac{k^i}{2}\frac{\pd^2}{\pd k^j\pd k^{m_1}}
	\right)P_\zeta(k)  \\
	&=& \frac{5H^2}{8\mpl^2 c_s \epsilon k^4}
	\left(
		 M_{jjm_1}(\hat q){\hat k}^{m_1}
		-3M_{m n m_1}(\hat q){\hat k}^{m}{\hat k}^{n}{\hat k}^{m_1}
	\right)\;.
\eea

\item {\it Quadratic consistency relation} ($n=2$): At this order, we have a new consistency relation
\begin{align}\label{CR n=2}
\lim_{\vec{q}\to 0}&M_{ijm_1m_2}(\hat{q})
\frac{\pd^2}{\pd q_{m_1}\pd q_{m_2}}
	\left(
	\frac{\la\gamma^{ij}_{\vec q}\zeta_{\vec k_1}\zeta_{\vec k_2}\ra'}{P_\gamma(q)}
	+\frac{\delta^{ij}}{3}\frac{\la\zeta_{\vec q}\zeta_{\vec k_1}\zeta_{\vec k_2}\ra'}{P_\zeta(q)}
	\right)
\nonumber\\
&\quad=
-
	M_{ijm_1m_2}(\hat{q})
	\left(
	\delta^{ij}\frac{\pd^2}{\pd k^{m_1}\pd k^{m_2}}
	+\frac{k^i}{3}\frac{\pd^3}{\pd k^j\pd k^{m_1}\pd k^{m_2}}
	\right)P_\zeta(k) \,.
\end{align}

From~(\ref{expand3pt}), the second derivatives of the scalar 3-point function is given by 
\beq
\lim_{\vec{q}\rightarrow 0}M_{i j m_1m_2}(\hat{q})\frac{\partial^2}{\partial q_{m_1}\partial q_{m_2}} 
\left(\frac{\delta_{ij}}{3}\frac{\la\zeta_{\vec q}\zeta_{\vec k_1}\zeta_{\vec k_2}\ra'}{P_\zeta(q)}
\right)\,
=  \frac{H^2}{\mpl^2}\,
\frac{5(1-c_s^2)}{24 c_s^3\epsilon k^5}M_{jjm_1m_2}(\hat{q}) \hat{k}^{m_1}\hat{k}^{m_2}\,,
\label{n=2LHS1}
\eeq
where we have again used the fact that terms with derivatives on $\Pi$ are projected out. Similarly, for the 3-point function
involving a soft tensor,
\bea
\nonumber
& & \lim_{\vec{q}\rightarrow 0} M_{i j m_1m_2}(\hat{q}) \frac{\partial^2}{\partial q_{m_1}\partial q_{m_2}}
\left( \frac{\la\gamma^{ij}_{\vec q}\zeta_{\vec k_1}\zeta_{\vec k_2}\ra'}{P_\gamma(q)}
\right) \\
& & = \frac{H^2}{\mpl^2}\,
\frac{5 \hat{k}^k\hat{k}^\ell}{8 c_s^3 \epsilon k_1^5}
\bigg( M_{k\ell jj}(\hat{q})(1+11c_s^2)
+14 c_s^2 M_{k \ell m_1m_2}(\hat{q})\hat{k}^{m_1}\hat{k}^{m_2}\bigg)\,.
\label{n=2LHS2}
\eeq
As advocated earlier, the $1/c_s^3$ contributions exactly cancel when combining~(\ref{n=2LHS1}) and~(\ref{n=2LHS2}), resulting in the left-hand side of~\eqref{CR n=2} being proportional to $1/c_s$:
\bea
\nonumber
\text{L.H.S.} &=&  \lim_{\vec{q} \to 0} M_{ijm_1m_2}(\hat{q})
\frac{\pd^2}{\pd q_{m_1}\pd q_{m_2}}
	\left(
	\frac{\la\gamma^{ij}_{\vec q}\zeta_{\vec k_1}\zeta_{\vec k_2}\ra'}{P_\gamma(q)}
	+\frac{\delta^{ij}}{3}\frac{\la\zeta_{\vec q}\zeta_{\vec k_1}\zeta_{\vec k_2}\ra'}{P_\zeta(q)}
	\right) \\
&=&  \frac{5 H^2}{4\mpl^2}\frac{\hat{k}^k\hat{k}^\ell}{c_s\epsilon k^5} \,
\bigg(6 M_{k\ell jj}(\hat{q})+7M_{k \ell m_1 m_2}(\hat{q})\hat{k}^{m_1}\hat{k}^{m_2}
\bigg)\,,
\eeq
This exactly matches the right-hand side of~(\ref{CR n=2}):
\bea
\nonumber
\text{R.H.S.}&=& -
	M_{ijm_1m_2}(\hat{q})
	\left(
	\delta^{ij}\frac{\pd^2}{\pd k^{m_1}\pd k^{m_2}}
	+\frac{k^i}{3}\frac{\pd^3}{\pd k^j\pd k^{m_1}\pd k^{m_2}}
	\right)P_\zeta(k) \\
	&=&   \frac{5 H^2}{4\mpl^2}\frac{\hat{k}^k\hat{k}^\ell}{c_s\epsilon k^5} \,
\bigg(6 M_{k\ell jj}(\hat{q})+7M_{k \ell m_1 m_2}(\hat{q})\hat{k}^{m_1}\hat{k}^{m_2}
\bigg)\,.
\eea
Note that the validity of the identity at $q^2$ order crucially relies on taking a particular linear combination of 3-point correlation functions. 

\item {\it Cubic consistency relation} ($n=3$): At $q^3$ order, the identity reads
\begin{align}\label{CR n=3}
\lim_{\vec{q}\to 0}&M_{ijm_1m_2 m_3}(\hat{q})
\frac{\pd^3}{\pd q_{m_1}\pd q_{m_2} \pd q_{m_3}}
	\left(
	\frac{\la\gamma^{ij}_{\vec q}\zeta_{\vec k_1}\zeta_{\vec k_2}\ra'}{P_\gamma(q)}
	+\frac{\delta^{ij}}{3}\frac{\la\zeta_{\vec q}\zeta_{\vec k_1}\zeta_{\vec k_2}\ra'}{P_\zeta(q)}
	\right)
\nonumber\\
&\quad=
-
	M_{ijm_1m_2 m_3}(\hat{q})\left(
	\delta^{ij}\frac{\pd^3}{\pd k^{m_1}\pd k^{m_2}\pd k^{m_3}}
	+\frac{k^i}{4}\frac{\pd^4}{\pd k^j\pd k^{m_1}\pd k^{m_2}\pd k^{m_3}}
	\right)P_\zeta(k) \,.
\end{align}
From~(\ref{expand3pt}), the $q^3$ contribution from $\langle \gamma\zeta\zeta\rangle$ is 
\bea
\nonumber
& &\lim_{\vec{q}\to 0} M_{ijm_1m_2 m_3}(\hat{q}) \frac{\pd^3}{\pd q_{m_1}\pd q_{m_2} \pd q_{m_3}}
	\frac{\la\gamma^{ij}_{\vec q}\zeta_{\vec k_1}\zeta_{\vec k_2}\ra'}{P_\gamma(q)}\\
& & =\frac{H^2}{\mpl^2}\frac{35}{16c_s^3\epsilon k^6}\hat{k}^k\hat{k}^\ell\hat{k}^{m_1}
\bigg( -27c_s^2M_{k\ell m_1m_2m_3}(\hat{q})  \hat{k}^{m_2}\hat{k}^{m_3}+ 
(1+5c_s^2)M_{jjk\ell m_1}(\hat{q})
\bigg)\,.
\label{n=3LHS1}
\eea
In deriving this, we have used the following properties of the projector (see Appendix~A):
\bea
\hat{q}_{m_1}M_{i j m_1mm}(\hat{q}) =0\;; \quad \hat{q}_{m_1}\hat{q}_{m_2}\hat{q}_{m_3}M_{i j m_1 m_2 m_3}(\hat{q})=0\;.
\eea

Similarly, the contribution from  $\langle \zeta\zeta\zeta\rangle$ is
\bea
\nonumber
& & \lim_{\vec{q}\to 0} M_{ijm_1m_2 m_3}(\hat{q}) \frac{\pd^3}{\pd q_{m_1}\pd q_{m_2} \pd q_{m_3}}
	\frac{\delta^{ij}}{3 P_\zeta(q)}\la\zeta_{\vec q}\zeta_{\vec k_1}\zeta_{\vec k_2}\ra'\\
	& & =-\frac{35H^2}{16\mpl^2}\frac{1-c_s^2}{c_s^3 k^6 \epsilon} M_{jj m_1 m_2 m_3}(\hat{q})\hat{k}^{m_1}\hat{k}^{m_2}\hat{k}^{m_3} \,.
\label{n=3LHS2}
\eeq
As before, the $1/c_s^3$ contributions in~(\ref{n=3LHS1}) and~(\ref{n=3LHS2}) cancel when taking the linear combination, resulting 
in the left-hand side of~\eqref{CR n=3} being proportional to $1/c_s$:
\bea
\nonumber
\text{L.H.S.} &=& \lim_{\vec{q}\to 0} M_{ijm_1m_2 m_3}(\hat{q})
\frac{\pd^3}{\pd q_{m_1}\pd q_{m_2} \pd q_{m_3}}
	\left(
	\frac{\la\gamma^{ij}_{\vec q}\zeta_{\vec k_1}\zeta_{\vec k_2}\ra'}{P_\gamma(q)}
	+\frac{\delta^{ij}}{3}\frac{\la\zeta_{\vec q}\zeta_{\vec k_1}\zeta_{\vec k_2}\ra'}{ P_\zeta(q)}
	\right) \\
&=& \frac{105H^2}{16\mpl^2}\frac{ \hat{k}^{m_1}\hat{k}^{m_2}\hat{k}^{m_3}}{c_s \ep k^6}
\bigg(2 M_{jjm_1m_2m_3}(\hat{q}) -9M_{k \ell m_1m_2m_3}(\hat{q}) \hat{k}^{k}\hat{k}^{\ell}
\bigg)\;.
\eeq
This agrees with the right-hand side of~(\ref{CR n=3}):
\bea
\nonumber
\text{R.H.S.} &=& -
	M_{ijm_1m_2 m_3}(\hat{q})
	\left(
	\delta^{ij}\frac{\pd^3}{\pd k^{m_1}\pd k^{m_2}\pd k^{m_3}}
	+\frac{k^i}{4}\frac{\pd^4}{\pd k^j\pd k^{m_1}\pd k^{m_2}\pd k^{m_3}}
	\right)P_\zeta(k) \\
	&=& \frac{105H^2}{16\mpl^2}\frac{ \hat{k}^{m_1}\hat{k}^{m_2}\hat{k}^{m_3}}{c_s \ep k^6}
\bigg(2 M_{jjm_1m_2m_3}(\hat{q}) -9M_{k \ell m_1m_2m_3}(\hat{q}) \hat{k}^{k}\hat{k}^{\ell}
\bigg)\;.
\eea

\end{itemize}

\section{Consistency Relations with One Scalar and One Tensor}
\label{zetagamma}

Next we consider the case where the hard modes consist of a scalar and a tensor, {\it i.e.}, $\mathcal{O}(\vec{k}_1,\vec{k}_2)= \gamma^{ab}_{\vec{k_1}} \zeta_{\vec{k_2}}$.
In this case, the identity~(\ref{wardalmosttheren>2}) reduces to
\begin{align}
&\lim_{\vec{q}\to 0} M_{j m_0 ... m_n}(\hat{q})
\frac{\pd^n}{\pd q^{m_1} ... q^{m_n}} 
\left(
\frac{ \langle\gamma^{j m_0}_{\vec q} \gamma^{a b}_{\vec k_1}\zeta_{\vec k_2}\rangle'}{P_\gamma(q)}
+\frac{\delta^{j m_0}}{3}\frac{\langle \zeta_{\vec q} \gamma^{a b}_{\vec k_1}\zeta_{\vec k_2}\rangle'}{P_\zeta(q)}
\right)
\nonumber\\
&=
M_{j m_0 ... m_n}(\hat{q})\Upsilon^{j m_0 c d}(\hat{k}_2)
\frac{\pd^n}
	{
	\pd k_2^{m_1}\dots \pd k_2^{m_n}
	}
	\la \gamma^{c d}_{\vec k_2}\gamma^{a b}_{\vec k_1}\ra\;,
\end{align}
where $\Upsilon$ was defined in~(\ref{Gamdef}). Note that the first and third lines of~(\ref{wardalmosttheren>2}), being proportional to $\langle \zeta\gamma\rangle$, dropped out. Only the $\Upsilon$ contribution survives. In checking these relations, we should consider (consistently) both left- and right-hand sides as a function of $\vec k_2$ and expand $\vec k_1 = -\vec{k}_2 - \vec{q}$ around $\vec{q} = 0$. This is related to the way of removing the momentum-conserving delta function from the consistency relations, when writing \eqref{wardalmosttheren>2} for primed correlation functions. See Appendix~B for a detailed treatment. 

Once again we expand the correlators in powers of $q$ up to cubic order, by first imposing the on-shell condition $\vec{k}_1 = -\vec{k}_2 - \vec{q}$:
\bea
\nonumber
\frac{ \langle\gamma^{j m_0}_{\vec q} \gamma^{a b}_{\vec k_1}\zeta_{\vec k_2}\rangle'}{P_\gamma(q)} &=&  
\frac{H^2}{\mpl^2 k^3 }\Pi_{jm_0 m n}(\hat q) 
\left(\Pi_{abmn}(\hat k)
+q^{l_1}\frac{\pd}{\pd k^{l_1}}\Pi_{abmn}(\hat k)
+\frac{1}{2}q^{l_1}q^{l_2}\frac{\pd^2}{\pd k^{l_1} \pd k_2}\Pi_{abmn}(\hat k)
+\dots\right) \nonumber\\
\nonumber
&\times&
\Bigg(
\frac{q}{k}\frac{\hat{k}\cdot \hat{q}}{8}
+\frac{q^2}{k^2}\frac{c_s^2+c_s+4 -3 c_s \left(c_s+1\right)
   (\hat{k}\cdot \hat{q})^2}{8 c_s \left(c_s+1\right)} 
\nonumber\\
\nonumber
&\quad& \quad
-\frac{q^3}{k^3}\frac{8+\left(9 c_s^3+18 c_s^2+17 c_s+16\right) \hat{k}\cdot \hat{q}
	-15c_s \left(c_s+1\right){}^2 (\hat{k}\cdot \hat{q})^3
	}{16 c_s \left(c_s+1\right){}^2}
+\dots
\Bigg) \,;
\\
\frac{\langle \zeta_{\vec q} \gamma^{a b}_{\vec k_1}\zeta_{\vec k_2}\rangle'}{P_\zeta(q)} &=& \frac{H^2}{\mpl^2 c_s (1+c_s) k^5}
\left(1+c_s+c_s^2
-\frac{q}{k}\frac{2+4c_s+6c_s^2+3c_s^3}{1+c_s}(\hat{q}\cdot \hat{k})
+\dots
\right) \nonumber\\
&\times&
\left(\Pi_{ijmn}(\hat k) q^m q^n
-\frac{1}{2}q^m q^{l_1} q^{l_2} k^n \frac{\pd^2}{\pd k^{l_1} \pd k^{l_2}} \Pi_{ijmn}(\hat k)  
+\dots
\right)\,.
\label{expand3ptzetagamma}
\eea
Notice that we have also Taylor expanded the polarization tensor $\Pi_{ijmn}(\hat k_1)=\Pi_{ijmn}({\hat k}_2+{\hat q})$ in powers of $q$, applying the identity 
\be\label{dPi}
\frac{\pd}{\pd k^{m_1}} \Pi_{i j m n}(\hat k)=-\frac{1}{k}
\left(
	\hat{k}_i \Pi_{m_1 j m n}(\hat k)
	+\hat{k}_j \Pi_{j m_1  m n}(\hat k)
	+\hat{k}_m \Pi_{i j m_1 n}(\hat k)
	+\hat{k}_n \Pi_{i j m m_1}(\hat k)
\right)\;
\ee
to simplify the last line of \eqref{expand3ptzetagamma}.
Since neither correlator gives at contribution at order $q^0$, the $n=0$ consistency relation is trivially satisfied.

\begin{itemize}

\item {\it Linear-gradient relation} ($n=1$): As seen from~(\ref{expand3ptzetagamma}), the expansion of $\langle\zeta\gamma\zeta\rangle/P_\zeta$ starts at order $q^2$ and hence does not contribute to the $n=1$ identity. The latter therefore reduces to
\be
\lim_{\vec{q}\to 0} M_{j m_0 m_1}(\hat{q})
\frac{\pd}{\pd q_{m_1}}
\frac{ \langle\gamma^{j m_0}_{\vec q} \gamma^{a b}_{\vec k_1}\zeta_{\vec k_2}\rangle'}{P_\gamma(q)}
=
	M_{j m_0 m_1}(\hat{q})
	\Upsilon^{j m_0 c d}(\hat{k}_2)
	\frac{\pd}{\pd k_2^{m_1}}	
\la \gamma^{c d}_{\vec k_2}\gamma^{a b}_{\vec k_1}\ra' \;.
\ee

The left-hand side gives
\bea
\nonumber
\text{L.H.S.}&=& \lim_{\vec{q}\to 0} M_{j m_0 m_1}(\hat{q})
\frac{\pd}{\pd q_{m_1}}
\frac{\langle\gamma^{j m_0}_{\vec q} \gamma^{a b}_{\vec k_1}\zeta_{\vec k_2}\rangle'}{P_\gamma(q)} \\
&=&   \frac{H^2}{4 \mpl^2 k_2^4}M_{m n m_1}(\hat{q})\Pi_{a b m n }(\hat k_2)\hat{k}_2^{m_1}\;,
\label{n=2 zeta gamma lhs}
\eea
Meanwhile, using the explicit expression for $\Upsilon$ given in~(\ref{Gamdef}), the right-hand side of the identity becomes
\begin{align}
\text{R.H.S.}&=
	M_{j m_0 m_1}(\hat{q})
	\left(
	\frac{1}{4}\delta^{j m_0}\hat{k}_2^c
	-\frac{1}{4}\delta^{j c}\hat{k}_2^{m_0}
	\right)
	\hat{k}_2^d
	\frac{\pd}{\pd k_2^{m_1}}	
	\left(
	\Pi_{c d a b}(\hat{k}_2) P_\gamma(k_2)
	\right)
\nonumber\\
& = M_{j m_1 m_0}(\hat{q})
\Pi_{a b j m_1 }(\hat{k}_2) 
\frac{H^2}{4\mpl^2 k_2^4} \hat{k}_2^{m_0}\;,
\end{align}
which agrees with~(\ref{n=2 zeta gamma lhs}).

\item {\it Quadratic consistency relation} ($n=2$): The identity at this order reads
\begin{align}
&\lim_{\vec{q}\to 0} M_{j m_0 m_1 m_2}(\hat{q})
\frac{\pd^2}{\pd q_{m_1}\pd q_{m_2}}
\left(
\frac{\langle\gamma^{j m_0}_{\vec q} \gamma^{a b}_{\vec k_1}\zeta_{\vec k_2}\rangle'}{P_\gamma(q)} 
+\frac{\delta^{j m_0}}{3}
\frac{\langle \zeta_{\vec q} \gamma^{a b}_{\vec k_1}\zeta_{\vec k_2}\rangle'}{P_\zeta(q)}
\right)
\nonumber\\
&=
	M_{j m_0 m_1 m_2}(\hat{q})
	\Upsilon^{j m_0 c d}(\hat{k}_2)
	\frac{\pd^2}{\pd k_2^{m_1}\pd k_2^{m_2}}	
\la \gamma^{c d}_{\vec k_2}\gamma^{a b}_{\vec k_1}\ra'\;.
\label{zetagamma n=2}
\end{align}

Using the differentiation properties of the projector $M$ and the polarization tensor $\Pi$, the first term on the left-hand side of the identity~(\ref{zetagamma n=2}) gives 
\bea
\nonumber
\text{L.H.S.~1st~term} &=& \lim_{\vec{q}\to 0}  M_{j m_0 m_1 m_2}(\hat{q})
\frac{\pd^2}{\pd q_{m_1}\pd q_{m_2}}
\frac{ \langle\gamma^{j m_0}_{\vec q} \gamma^{a b}_{\vec k_1}\zeta_{\vec k_2}\rangle'}{P_\gamma(q)}\\
\nonumber
&=& 
\frac{H^2}{2\mpl^2 k_2^5}\Bigg\{
\Pi_{ a b m n}(\vec{k}_2)
\bigg(
	\frac{4+c_s+c_s^2}
	{ c_s (1+c_s)}M_{m n m_1 m_1}(\hat{q})
-4 M_{m n m_1 m_2}(\hat{q})\hat{k}_2^{m_1}\hat{k}_2^{m_2} \\
\nonumber
& &~~~~~~~~~~ -M_{ m_1 m_2 m n}(\hat{q})\hat{k}_2^{m_1}\hat{k}_2^{m_2}\Bigg) 
\nonumber\\
&\quad& ~~~~~~~~~~ -M_{ m n m_1 m_2}(\hat q)\hat{k}_2^{m_2}
\bigg(
\Pi_{m n m_1 a}(\hat{k}_2)\hat{k}_2^b
+\Pi_{m n m_1 b}(\hat{k}_2) \hat{k}_2^a
\bigg)\Bigg\} \,,
\label{zetagamman=2LHS1}
\eea
where we have used the fact that
\be
M_{j m_0 m_1 m_2}(\hat{q})\Pi_{j m_0 m n}(\hat{q})=M_{m n m_1 m_2}(\hat{q})+M_{n m m_1 m_2}(\hat{q})-\frac{2}{3}\delta_{m n}M_{j j m_1 m_2}(\hat{q})\;.
\label{otheridentity}
\ee
Meanwhile, the second term on the left-hand side of~(\ref{zetagamma n=2}) gives
\bea
\nonumber
\text{L.H.S.~2nd~term} &=& \lim_{\vec{q}\to 0}  M_{j m_0 m_1 m_2}(\hat{q}) 
\frac{\pd^2}{\pd q_{m_1}\pd q_{m_2}}\left(\frac{\delta^{j m_0}}{3}  \frac{\langle \zeta_{\vec q} \gamma^{a b}_{\vec k_1}\zeta_{\vec k_2}\rangle'}{P_\zeta(q)} \right)\\
&= &
\frac{2H^2}{3\mpl^2 k_2^5} \frac{1+c_s+c_s^2}{ c_s (1+c_s)}  M_{jjmn}(\hat{q})
\Pi_{a b m n}(\hat{k}_2)\;.
\label{zetagamman=2LHS2}
\eea

The complicated $c_s$ dependence in~(\ref{zetagamman=2LHS1}) and~(\ref{zetagamman=2LHS2}) simplifies tremendously when combining these terms, leaving us with
\bea
\nonumber
{\rm L.H.S.}
&=& \frac{H^2}{2\mpl^2 k_2^5} \Bigg\{
\Pi_{abmn}(\hat{k}_2)
\left(
	M_{jjmn}(\hat{q})
	-4\hat{k}_2^{m_1}\hat{k}_2^{m_2}M_{m n m_1 m_2}(\hat{q})
	-\hat{k}_2^{m_1}\hat{k}_2^{m_2}M_{ m_1 m_2 m n}(\hat{q})
\right)\nonumber\\
& & ~~~~~~~~~~~~ - M_{ m n m_1 m_2}(\hat q)\hat{k}_2^{m_2}
\left(
\Pi_{m n m_1 a}(\hat{k}_2)\hat{k}_2^b
+\Pi_{m n m_1 b}(\hat{k}_2) \hat{k}_2^a
\right) \Bigg\}\,.
\label{zetagamman=2LHS}
\eea

On the other hand, substituting for $\Upsilon$, the right-hand side of the identity~(\ref{zetagamma n=2}) becomes
\bea
\nonumber
{\rm R.H.S.}&=&	M_{j m_0 m_1 m_2}(\hat{q})
	\left(
	\frac{1}{4}\delta^{j m_0}\hat{k}_2^c\hat{k}_2^d
	-\frac{1}{4}\delta^{j c}\hat{k}_2^{m_0}\hat{k}_2^d
	\right)
	\frac{\pd^2}{\pd k_2^{m_1}\pd k_2^{m_2}}	
	\left(
	\Pi_{c d a b}(\hat{k}_2) P_\gamma(k_2)
	\right)\nonumber\\
&=&\frac{H^2}{\mpl^2 k_2^5}M_{j m_0 m_1 m_2}(\hat q)
\Bigg[-2\Pi_{j m_0 a b}(\hat k_2)\hat{k}_2^{m_1}\hat{k}_2^{m_2}
	+\frac{1}{2}\Pi_{m_1 m_2 a b}(\hat k_2)\delta_{j m_0}
\nonumber\\
&&	-\frac{1}{2}\Pi_{m_1 m_2 a b}(\hat k_2)\hat{k}_2^{j}\hat{k}_2^{m_0}
-\frac{1}{2}\Pi_{j m_0 m_1 b}(\hat k_2)\hat{k}_2^{a}\hat{k}_2^{m_2}
-\frac{1}{2}\Pi_{j m_0 m_1 a}(\hat k_2)\hat{k}_2^{b}\hat{k}_2^{m_2}
\Bigg] \,.
\label{zetagamman=2RHS}
\eea
Here, we have repeatedly applied \eqref{dPi} to obtain the second equality.
One finds that~(\ref{zetagamman=2LHS}) and~(\ref{zetagamman=2RHS}) match perfectly --- the $n=2$ identity checks out!

\item {\it Cubic consistency relation} ($n=3$): The identity at this order is
\begin{align}
&\lim_{\vec{q}\to 0} M_{j m_0 m_1 m_2 m_3}(\hat{q})
\frac{\pd^3}{\pd q_{m_1}\pd q_{m_2}\pd q_{m_3}}
\left(
\frac{ \langle\gamma^{j m_0}_{\vec q} \gamma^{a b}_{\vec k_1}\zeta_{\vec k_2}\rangle'}{P_\gamma(q)}
+\frac{\delta^{j m_0}}{3}
\frac{\langle \zeta_{\vec q} \gamma^{a b}_{\vec k_1}\zeta_{\vec k_2}\rangle'}{ P_\zeta(q)}
\right)
\nonumber\\
&=
	M_{j m_0 m_1 m_2 m_3}(\hat{q})
	\Upsilon^{j m_0 c d}(\hat{k}_2)
	\frac{\pd^3}{\pd k_2^{m_1}\pd k_2^{m_2} \pd k_3^{m_3}}	
\la \gamma^{c d}_{\vec k_2}\gamma^{a b}_{\vec k_1}\ra'\;.
\end{align}

The first term on the left-hand side becomes 
\bea
\nonumber
\text{L.H.S.~1st~term} &=&  \lim_{\vec{q}\to 0} M_{j m_0 m_1 m_2m_3}(\hat{q})
\frac{\pd^3}{\pd q_{m_1}\pd q_{m_2}\pd q_{m_3}}
\frac{\langle\gamma^{j m_0}_{\vec q} \gamma^{a b}_{\vec k_1}\zeta_{\vec k_2}\rangle'}{P_\gamma(q)} \\
\nonumber
&=&\frac{3H^2}{4\mpl^2 k_2^6}
M_{m n m_1 m_2 m_3}(\hat q)
\Bigg\{
\Pi_{a b m n }(\hat{k}_2)
\bigg(
-\frac{16+17c_s+18c_s^2+9c_s^3}{c_s(1+c_s)^2}\hat{k}_2^{m_1}\delta^{m_2 m_3} \\
\nonumber
& & ~~~~~~~~~~~~~~~~~~~~~~~~~~~~~~~~~~~~~~~~~~~~~~~~ +15\hat{k}_2^{m_1}\hat{k}_2^{m_2}\hat{k}_2^{m_3}\bigg) \\
\nonumber
&& ~~~~~~~~~~~~~~~~~~~~~~~~~~~~~+ 2k_2 \frac{\pd}{\pd k_2^{m_1}}\Pi_{a b m n}(\hat{k}_2)
	\left(
	\frac{4+c_s+c_s^2}{c_s(1+c_s)}\delta^{m_2 m_3}-3\hat{k}_2^{m_2}\hat{k}_2^{m_3}
	\right) \\
& & ~~~~~~~~~~~~~~~~~~~~~~~~~~~~~ +k_2^2\frac{\pd^2}{\pd k_2^{m_1}\pd k_2^{m_2}}\Pi_{a b m n}(\hat{k}_2)\hat{k}_2^{m_3}
\Bigg]\;.
\label{zetagamman=3LHS1}
\eea
In deriving this expression we have used the properties of $M$ given in~\eqref{P Prop 1})$-$\eqref{P Prop 4} of Appendix~A to discard terms involving $M_{m n m_1 m_2 m_3}(\hat q) {\hat q}^{m_1}{\hat q}^{m_2}{\hat q}^{m_2}$ and $M_{m n j j m_1}(\hat q){\hat q}^{m_3}$. Similarly, the second term on the left-hand side gives
\bea
\nonumber
\text{L.H.S.~2nd~term} &=& \lim_{\vec{q}\to 0} M_{j m_0 m_1 m_2 m_3}(\hat{q})
\frac{\pd^3}{\pd q_{m_1}\pd q_{m_2}\pd q_{m_3}}
\left( \frac{\delta^{j m_0}}{3}
\frac{\langle \zeta_{\vec q} \gamma^{a b}_{\vec k_1}\zeta_{\vec k_2}\rangle'}{ P_\zeta(q)} \right) \\
\nonumber
&=& \frac{6H^2}{\mpl^2 k_2^6}M_{m n m_1 j j }(\hat q)
	\Bigg(
	\frac{2+4c_s+6c_s^2+3c_s^3}{c_s(1+c_s)^2}{\hat k}_2^{m_1}\Pi_{a b m n}(\hat k_2) \\
& &~~~~~~~~~~~~~~~~~~~~~~~~~~  	-\frac{1+c_s+c_s^2}{c_s (1+c_s)}k_2 \frac{\pd}{\pd k_2^{m_1}}\Pi_{a b m n}(\hat k_2)
	\Bigg)\;.
\label{zetagamman=3LHS2}
\eea
Once again, the complicated $c_s$ dependence simplifies dramatically when combining these two terms, leaving us with 
\bea
\nonumber
{\rm L.H.S.} &=& \lim_{\vec{q}\to 0} M_{j m_0 m_1 m_2 m_3}(\hat{q})
\frac{\pd^3}{\pd q_{m_1}\pd q_{m_2}\pd q_{m_3}}
\left(
\frac{ \langle\gamma^{j m_0}_{\vec q} \gamma^{a b}_{\vec k_1}\zeta_{\vec k_2}\rangle'}{P_\gamma(q)}
+\frac{\delta^{j m_0}}{3}
\frac{\langle \zeta_{\vec q} \gamma^{a b}_{\vec k_1}\zeta_{\vec k_2}\rangle'}{ P_\zeta(q)}
\right) \\
\nonumber
&=& \frac{3H^2}{4\mpl^2 k_2^6}
M_{m n m_1 m_2 m_3}(\hat q)
\Bigg[
\Pi_{a b m n }(\hat k_2)
\left(
15\hat{k}_2^{m_1}\delta^{m_2 m_3}
+15\hat{k}_2^{m_1}\hat{k}_2^{m_2}\hat{k}_2^{m_3}\right)
\nonumber\\
&+& k_2 \frac{\pd}{\pd k_2^{m_1}}\Pi_{a b m n}(\hat k_2)
	\left(
	-6\delta^{m_2 m_3}-6\hat{k}_2^{m_2}\hat{k}_2^{m_3}
	\right)
+k_2^2\frac{\pd^2}{\pd k_2^{m_1}\pd k_2^{m_2}}\Pi_{a b m n}(\hat k_2)\hat{k}_2^{m_3}
\Bigg]\;.
\eea
By explicitly substituting for $\Pi$ and using~(\ref{dPi}), we have checked with Mathematica that the result agrees with the right-hand side --- the $n=3$ identity checks out!

\end{itemize}

\section{Consistency Relations with Two Tensors}
\label{gammagamma}

Finally we consider the case where the hard modes consist of two tensors, {\it i.e.}, $\mathcal{O}(\vec{k}_1,\vec{k}_2)=\gamma^{a b}_{\vec k_1}\gamma^{c d}_{\vec k_2}$.
In this case, the identity~(\ref{wardalmosttheren>2}) reduces to \footnote{We have used the fact that $\Gamma_{jm_0 ab mn}(-\vec k)=\Gamma_{jm_0 a b m n}(\vec k)$, so that the derivative on $\Gamma$ with respect to $\vec k_2$ in (\ref{wardalmosttheren>2}) is now translated to derivative w.~r.~t.~the hard momentum $\vec k_1 =\vec k$. }
\begin{align}
\lim_{\vec{q}\to 0}& M_{j m_0 ... m_n}(\hat{q})
\frac{\pd^n}{\pd q^{m_1} ... q^{m_n}} 
\left(
\frac{\langle\gamma^{j m_0}_{\vec q} \gamma^{a  b}_{\vec k_1}\gamma^{ c d}_{\vec k_2}\rangle'}{P_\gamma(q)} 
+\frac{\delta^{j m_0}}{3}
\frac{\langle \zeta_{\vec q} \gamma^{a b}_{\vec k_1}\gamma^{cd}_{\vec k_2}\rangle'}{P_\zeta(q)}
\right)
\nonumber\\
&=
-M_{j m_0 ... m_n}(\hat{q})
\left(
	\delta_{j m_0}
	\frac{\pd^n}
	{
	\pd k^{m_1}\dots \pd k^{m_n}
	}
	-\frac{\delta_{n0}}{2}\delta_{j m_0}
	+\frac{k^j}{n+1}
	\frac{\pd^{n+1}}
	{\pd k^{m_0}\dots \pd k^{m_n}}
\right)
	\la \gamma^{ab}_{\vec k}\gamma^{cd}_{-\vec k}
	\ra'
\nonumber\\
&\quad+
	M_{j m_0 ... m_n}(\hat{q})
\left(
	\Gamma_{j m_0 a b m n}(\hat{k})
\frac{\pd^n \la \gamma^{m n}_{\vec k}\gamma^{c d}_{-\vec k}\ra'}
	{
	\pd k^{m_1}\dots \pd k^{m_n}
	}
+
\frac{\pd^n \Gamma_{j m_0 c d m n}(\hat{k})}
	{
	\pd k^{m_1}\dots \pd k^{m_n}
	}
	\la \gamma^{m n}_{\vec k}\gamma^{a b}_{-\vec k}\ra'
\right)\;,
\end{align}
where $\Gamma$ was defined in~(\ref{Gamdef}). On the left-hand side, an important simplification occurs when noting that $\langle\zeta\gamma\gamma\rangle/P_\zeta$ is subleading in $\epsilon$ compared to $\langle\gamma\gamma\gamma\rangle/P_\gamma$. In other words, only the latter contributes to leading order in slow-roll.

In order to obtain the soft limit expansion of the correlation function, we impose the on-shell condition $\vec{k_2}=-\vec{k_1}-\vec{q}$ and Taylor-expand in $q$:
\beq
\frac{\langle\gamma^{ij}_{\vec q} \gamma^{k\ell}_{\vec k_1}\gamma^{mn}_{\vec k_2}\rangle'}{P_\gamma(q)}&=&\frac{2H^2}{4\mpl^2k^2}\left( 1-\frac{5\hat{k}\cdot \vec{q}}{2 k}-\frac{5\left( -7(\hat{k}\cdot\vec{q})^2+q^2 \right)}{6 k^2}-\frac{105 (\hat{k}\cdot \vec{q})^3-35q^2\hat{k}\cdot\vec{q}+q^3}{8k^3}+\ldots \right)\nonumber \\
&\times& \Pi^{ij}_{\;~aa'}(\hat{q})\Pi^{k\ell}_{\;~bb'}(\hat{k})\left( k^a\delta_{bc}-(k+q)^b\delta_{ac}+q^c\delta_{ab} \right) \left( k^{a'}\delta_{b'c'}-(k+q)^{b'}\delta_{a'c'}+q^{c'}\delta_{a'b'} \right)\nonumber \\
&\times&\left(\Pi^{mn}_{\;~cc'}(\hat{k})+q^{\ell_1}\frac{\pd}{\pd k^{\ell_1}}\Pi^{mn}_{\;~cc'}(\hat{k})+\frac{1}{2} q^{\ell_1}q^{\ell_2}\frac{\pd^2}{\pd k^{\ell_1} \pd k^{\ell_2}}\Pi^{mn}_{\;~cc'}(\hat{k})\right. \nonumber \\
&&\quad \left. +\frac{1}{6} q^{\ell_1}q^{\ell_2}q^{\ell_3}\frac{\pd^3}{\pd k^{\ell_1} \pd k^{\ell_2}\pd k^{\ell_2}}\Pi^{mn}_{\;~cc'}(\hat{k})+\ldots\right)
\,.
\eeq

\begin{itemize}

\item {\it Lowest-order relation} ($n=0$):  The identity in this case reduces to
\begin{align}
\lim_{\vec{q} \to 0}& M_{j m_0}(\hat{q})
\frac{\langle\gamma^{j m_0}_{\vec q} \gamma^{a  b}_{\vec k_1}\gamma^{ c d}_{\vec k_2}\rangle'}{P_\gamma(q)} 
=
-M_{j m_0}(\hat{q})
\left(
	\delta_{j m_0}
	+k^j
	\frac{\pd}
	{\pd k^{m_0}}
\right)
	\la \gamma^{ab}_{\vec k}\gamma^{cd}_{-\vec k}
	\ra'
\nonumber\\
&\quad+
	M_{j m_0}(\hat{q})
\left(
	\Gamma_{j m_0 a b m n}(\hat{k})
	\la \gamma^{m n}_{\vec k}\gamma^{c d}_{-\vec k}\ra'+
	\Gamma_{j m_0 c d m n}(\hat{k})
	\la \gamma^{m n}_{\vec k}\gamma^{a b}_{-\vec k}\ra'
\right)
	\;.
\end{align}
For dilation, $M_{j m_0}\propto \delta_{j m_0}$, the relation is trivially satisfied.  For anisotropic scaling,\footnote{In this case, we can assume that $M_{j m_0}$ is symmetric, traceless and transverse to $q$, without loss of generality.} we find that 
\bea
\nonumber
\text{L.H.S.~}&=&\lim_{\vec{q} \to 0} M_{j m_0}(\hat{q})
\frac{\langle\gamma^{j m_0}_{\vec q} \gamma^{a  b}_{\vec k_1}\gamma^{ c d}_{\vec k_2}\rangle'}{P_\gamma(q)}  \\
&=& \frac{3 H^2}{\mpl^2 k^3} M_{m n}(\hat{q})\hat{k}^{m}\hat{k}^{n}\Pi_{abcd}(\hat k)\;.
\eea
With the aid of~\eqref{dPi} and~\eqref{Gamdef}, it is straightforward to show that this matches the right-hand side of the identity.

\item {\it Linear-gradient relation} ($n=1$): The $n=1$ relation reads
\begin{align}\label{gammagamma1}
\lim_{q\to 0}& M_{j m_0 m_1}(\hat{q})\frac{\pd}{\pd q^{m_1}}
\left(
\frac{1}{P_\gamma(q)} 
\langle\gamma^{j m_0}_{\vec q} \gamma^{a  b}_{\vec k_1}\gamma^{ c d}_{\vec k_2}\rangle'
\right)
=
-M_{j m_0 m_1}(\hat{q})
\left(
	\delta_{j m_0} \frac{\pd }{\pd k_1^{m_1}}
	+\frac{k_1^j}{2}
	\frac{\pd^2}
	{\pd k_1^{m_0}\pd k_1^{m_1}}
\right)
	\la \gamma^{ab}_{\vec k_1}\gamma^{cd}_{-\vec k_1}
	\ra'
\nonumber\\
&\quad+
	M_{j m_0 m_1}(\hat{q})
\left(
	\Gamma_{j m_0 a b m n}(\hat{k}_1)
	\frac{\pd \la \gamma^{m n}_{\vec k_1}\gamma^{c d}_{-\vec k_1}\ra'}{\pd k_1^{m_1}}+
	\frac{\pd \Gamma_{j m_0 c d m n}(\hat{k}_1)}{\pd k_1^{m_1}}
	\la \gamma^{m n}_{\vec k_1}\gamma^{a b}_{-\vec k_1}\ra'
\right)
	\;.
\end{align}
As a result of a lengthy computation, we find that the left-hand side becomes 
\bea
\text{L.H.S.}&=&M_{m n \ell}(\hat{q})\frac{H^2}{\mpl^2 k^4}
\Bigg[-3\hat{k}^m \hat{k}^n \Pi_{a b \ell c}(\hat k)\hat{k}^{d}-3\hat{k}^m \hat{k}^n \Pi_{a b \ell d}(\hat k)\hat{k}^{c}
-\frac{15}{2}\hat{k}^m \hat{k}^n\hat{k}^\ell \Pi_{a b c d}(\hat k)\nonumber\\
& &~~~~~~~~~~~~~~~~~~~~ +\frac{3}{2}\hat{k}^\ell \Pi_{a b m m_1}(\hat k) \Pi_{c d n m_1}(\hat k)
-\frac{3}{2}\hat{k}^\ell \Pi_{c d m m_1}(\hat k) \Pi_{a b n m_1}(\hat k)\nonumber \\
& & ~~~~~~~~~~~~~~~~~~~~ + \delta_{mn}\left(\frac{5}{2}\hat{k}^\ell  \Pi_{a b c d}(\hat k) +\hat{k}^d  \Pi_{a b c \ell}(\hat k) +\hat{k}^c  \Pi_{a b d \ell} (\hat k) \right)
\Bigg]\;.
\eea
By plugging the explicit form for $\Pi$ and its derivatives, it is straightforward, though tedious, to show using Mathematica that this expression matches the right-hand side of~\eqref{gammagamma1}.

\item {\it Quadratic consistency relation} ($n=2$): The identity at this order reduces to
\begin{align}
\lim_{q\to 0}& M_{j m_0 m_1 m_2}(\hat{q})\frac{\pd^2}{\pd q^{m_1}\pd q^{m_2}}
\left(
\frac{1}{P_\gamma(q)} 
\langle\gamma^{j m_0}_{\vec q} \gamma^{a  b}_{\vec k_1}\gamma^{ c d}_{\vec k_2}\rangle'
\right)
\nonumber\\
&=
-M_{j m_0 m_1 m_2}(\hat{q})
\left(
	\delta_{j m_0}\frac{\pd^2}{\pd k_1^{m_1}\pd k_1^{m_2}}
	+\frac{k_1^j}{3}
	\frac{\pd^3}
	{\pd k_1^{m_0}\pd k_1^{m_1}\pd k_1^{m_2}}
\right)
	\la \gamma^{ab}_{\vec k_1}\gamma^{cd}_{-\vec k_1}
	\ra'
\nonumber\\
&\quad+
	M_{j m_0 m_1 m_2}(\hat{q})
\left(
	\Gamma_{j m_0 a b m n}(\hat{k}_1)
	\frac{\pd^2 \la \gamma^{m n}_{\vec k_1}\gamma^{c d}_{-\vec k_1}\ra'}{\pd k_1^{m_1}\pd k_1^{m_2}}+
	\frac{\pd^2 \Gamma_{j m_0 c d m n}(\hat{k}_1)}{\pd k_1^{m_1}\pd k_1^{m_2}}
	\la \gamma^{m n}_{\vec k_1}\gamma^{a b}_{-\vec k_1}\ra'
\right)
	\;.
\end{align}

Again, the left hand side can be written in the following form
\beq
\text{L.H.S.}&=&\frac{H^2}{\mpl^2 k^5}M_{j m_0 m_1m_2}(\hat{q})\Bigg[ -10 \delta_{jm_0}\Pi_{abcd}(\hat{k})\hat{k}^{m_1}\hat{k}^{m_2}+35 \Pi_{abcd}(\hat{k})\hat{k}^{m_0}\hat{k}^{m_1}\hat{k}^{m_2} \nonumber \\
&&+2 \delta_{jm_0}\Pi_{abm_2s}(\hat{k})\Pi_{cdm_1s}(\hat{k})-\frac{15}{2} \Pi_{abjs}(\hat{k})\Pi_{cdm_1s}(\hat{k})\hat{k}^{m_0}\hat{k}^{m_2}\nonumber \\
&&+\frac{15}{2} \Pi_{abm_0s}(\hat{k})\Pi_{cdm_1s}(\hat{k})\hat{k}^{j}\hat{k}^{m_2}-3k \hat{k}^{m_2}\Pi_{abm_1s_1}(\hat{k})\frac{\pd}{\pd k^{m_0}}\Pi_{cdjs_1}(\hat{k})\nonumber \\
&&+3k \Pi_{abm_1s_1}(\hat{k})\frac{\pd}{\pd k^{j}}\Pi_{cdm_0s_1}(\hat{k})\hat{k}^{m_2} +\frac{5}{2} \delta_{jm_0} \Pi_{abss_1}(\hat{k})\frac{\pd}{\pd k^{m_2}}\Pi_{cd s s_1}(\hat{k})\hat{k}^{m_1}\nonumber \\
&&-\frac{15k}{2}\Pi_{abm_1s_1}(\hat{k})\frac{\pd}{\pd k^{m_0}}\Pi_{cd s s_1}(\hat{k})\hat{k}^{m_1}\hat{k}^{j}\hat{k}^{m_2}-\frac{k^2}{2} \delta_{jm_0} \Pi_{abss_1}(\hat{k})\frac{\pd^2}{\pd k^{m_1}\pd k^{m_2}}\Pi_{cd s s_1}(\hat{k})\nonumber \\
&&+\frac{3k^2}{2}  \Pi_{abss_1}(\hat{k})\frac{\pd^2}{\pd k^{m_1}\pd k^{m_2}}\Pi_{cd s s_1}(\hat{k})\hat{k}^j\hat{k}^{m_0} \Bigg]\,.
\eeq
Despite its complexity, explicit substitution in Mathematica, taking into account the explicit form of $\Pi$ and its derivatives, shows that this relation matches the right-hand side of the identity.

\item {\it Cubic consistency relation} ($n=3$): At this order, expressions become quite cumbersome. The cubic contribution to the 3-point function is given by
\beq
\frac{\langle\gamma^{ij}_{\vec q} \gamma^{k\ell}_{\vec k_1}\gamma^{mn}_{\vec k_2}\rangle'}{P_\gamma(q)}\supset \frac{3H^2}{4\mpl^2k^5}\Pi_{ijaa'}(\hat{q})\Pi_{k\ell bb'}(\hat{k})(\hat{k})\Bigg[ \frac{1}{2}q^{\ell_1}q^{\ell^2}(-q^{b'}\delta_{a'c'}+q^{c'}\delta_{a'b'})\frac{\pd^2}{\pd k^{\ell_1}\pd k^{\ell_2}} \Pi_{mnbc'}(\hat{k})k^a\nonumber \\
+\frac{1}{6}q^{\ell_1}q^{\ell_2}q^{\ell_3}\frac{\pd^2}{\pd k^{\ell_1}\pd k^{\ell_2}\pd k^{\ell_3}} \Pi_{mnbb'}(\hat{k})k^ak^{a'}-\frac{105(\vec{k}\cdot\vec{q})^3-35k^2q^2\vec{k}\cdot\vec{q}+2k^3q^3}{8k^6}\Pi_{mnbb'}\nonumber \\
-\frac{5\left( -7(\vec{k}\cdot\vec{q})^2+k^2q^2 \right)}{6 k^4}\Big( 2\Pi_{mnbc'}(\hat k) k^a \left(-q^{b'}\delta_{a'c'}+q^{c'}\delta_{a'b'}\right)+q^{\ell_1}\frac{\pd}{\pd k^{\ell_1}} \Pi_{mnbb'}(\hat{k})k^ak^{a'}\Big)\nonumber \\
-\frac{5\vec{k}\cdot\vec{q}}{2k^2}\Big( \Pi_{mncc'}(\hat{k})\left(-q^b\delta_{ac}+q^{c}\delta_{ab}\right)\left(-q^{b'}\delta_{a'c'}+q^{c'}\delta_{a'b'}\right) +2q^{\ell_1}\left(-q^{b'}\delta_{a'c'}+q^{c'}\delta_{a'b'}\right)\frac{\pd}{\pd k^{\ell_1}} \Pi_{mnbc'}(\hat{k})k^a\nonumber \\
+\frac{1}{2}q^{\ell_1}q^{\ell_2}\frac{\pd^2}{\pd k^{\ell_1}\pd k^{\ell_2}} \Pi_{mnbb'}(\hat{k})k^{a}k^{a'} \Big)
\Bigg]\,.\nonumber
\eeq
After differentiating this expression with respect to $q$, the tedious calculation using Mathematica shows that the consistency condition holds at this order as well.

\end{itemize}

\section{Conclusions}
\label{conclusec}

Single-field primordial perturbations satisfy an infinite number of consistency relations constraining at each $n\geq 0$ the $q^n$ behavior of
correlation functions in the soft limit $\vec{q}\rightarrow 0$. These are the consequence of Ward identities for an infinite set of residual, global
symmetries, which are non-linearly realized on the perturbations~\cite{Hinterbichler:2013dpa}. Equivalently, they are also the consequence of the Slavnov-Taylor identity
for spatial diffeomorphisms~\cite{Berezhiani:2013ewa}. 

In this paper, we performed a number of non-trivial checks of the consistency relations up to and including $q^3$ order, focusing for simplicity on identities
involving 3-point functions with a soft external leg. We considered all possible scalar and tensor combinations of hard modes, {\it i.e.}, $\zeta\zeta$, $\zeta\gamma$, and $\gamma\gamma$.
Our computation was done in the context of single-field, slow-roll inflation with arbitrary constant sound speed $c_s$. For this purpose, the 3-point functions $\langle \gamma\zeta\zeta\rangle$
and $\langle \zeta\gamma\gamma \rangle$ with $c_s \neq 1$ were computed here for the first time. 

The $n=2$ and $n=3$ checks are particularly non-trivial and physically interesting. Indeed, at order $q^2$ and higher, part of the information encoded in
correlation functions is physical --- it represents spatial curvature, which is model-dependent and therefore cannot be constrained in terms of
lower-point functions. Nevertheless, certain linear combinations of soft correlators are constrained by consistency relations~\cite{Hinterbichler:2013dpa}.
This showed up explicitly in our checks through non-trivial $c_s$ dependence --- only by taking certain linear combinations dictated by the identities did
the complicated $c_s$ dependence of the 3-point functions simplify to match that of the 2-point functions. 

The $n\geq 2$ checks are important because the corresponding identities have not yet been derived through standard
background-wave arguments. In particular, the mode functions in momentum-space
acquire an imaginary part at $n=3$, and hence cannot fully describe a `classical' background wave. Our
checks nevertheless confirm that part of the $q^3$ information can be thought of as classical, and is constrained in
a background-wave manner.

It is clear that soft limits of cosmological correlation functions encode a wealth of information about the primordial universe. A number of interesting future
avenues naturally suggest themselves:

\begin{itemize}

\item The lowest-order ($n=0,1$) identities can be violated whenever their underlying assumptions --- single field, attractor background, adiabatic vacuum --- 
are not satisfied. In this sense, the identities are not tautological: they are physical statements that can be tested observationally.
We expect the same holds true for the higher-order ($n\geq 2$) identities, but it would be reassuring to seek explicit violations, {\it e.g.}, in multi-field scenarios or with
non-adiabatic vacua. This is currently in progress~\cite{usfuture}. 

\item As is well-known from pion physics~\cite{Weinberg:1966kf}, another probe of higher-$q$ dependence is to consider multiple soft limits. This is currently under investigation
for the conformal algebra associated with scalar perturbations~\cite{Marcofuture}. It would be enlightening to generalize and include tensor modes as well. 

\item The higher-order identities were discovered through formal, field-theoretic methods~\cite{Hinterbichler:2013dpa,Goldberger:2013rsa,Berezhiani:2013ewa,Assassi:2012zq}. 
A more physical derivation of the consistency relations using background-wave arguments would undoubtedly offer new insights on their physical origin. 

\end{itemize}

{\bf Acknowledgements:} We would like to thank Paolo Creminelli and Leonardo Senatore for useful discussions. This work is supported in part by NASA ATP grant NNX11AI95G, NSF CAREER Award PHY-1145525 (J.K.) and funds from the University of Pennsylvania (L.B.).  J.W. is supported by the National Science Foundation grant PHY-1316665 and PHY-1214000. 
J.W. thanks the Center for Particle Cosmology at U. Penn for hospitality while this work was completed.

\appendix

\section*{Appendix A: Properties of the Projectors}
\label{Property of P}
\renewcommand{\theequation}{A-\Roman{equation}}
\setcounter{equation}{0} 

In this Appendix we present some useful relations for the projection operator $M_{i l_0 \dots l_n}$, which follow directly from its defining properties. 
By definition the projector satisfies the following transversality condition
\be\label{Trans P}
\hat{q}_i\left(
M_{ij m_1 m_2 m_3}(\hat{q})+M_{j i m_1 m_2 m_3}(\hat{q})
-\frac{2}{3}\delta_{ij}M_{kk m_1 m_2 m_3}(\hat{q})
\right)
=0\;.
\ee
Contracting~\eqref{Trans P} with $\delta_{m_1 m_2}$,\footnote{Notice that we do not get anything new if we contract~\eqref{Trans P} with $\delta_{j m_1}$.}
we get 
\be
\hat{q}_i M_{ij m m m_3}(\hat{q})+\hat{q}_i M_{j i m m m_3}(\hat{q})=0\;.
\ee
Given that the projector obeys the trace condition~(\ref{Mcond}), we arrive at 
\be\label{P Prop 1}
\hat{q}_i M_{ij k m m}(\hat{q})=\hat{q}_i M_{j k i m m }(\hat{q})=0\;.
\ee

By contracting both sides of~\eqref{Trans P} with $\hat{q}_j$, we get 
\be\label{P Prop 2}
\hat{q}_i\hat{q}_j M_{ij m_1 m_2 m_3}(\hat{q})=\frac{1}{3}M_{kk m_1 m_2 m_3}(\hat{q})\;.
\ee
Also we can multiply~\eqref{Trans P} by $\hat{q}_{m_1}$.  With the aid of~\eqref{P Prop 2}, this leads us to obtain
\be\label{P Prop 3}
\frac{1}{3} M_{k k j m_2 m_3}(\hat{q})
+\hat{q}_i\hat{q}_{m_1} M_{j m_2 m_3 i m_1}(\hat{q})
-\frac{2}{3}\hat{q}_j \hat{q}_{m_1} M_{k k m_2 m_3 m_1}(\hat{q})
=0\;.
\ee
Combining~\eqref{P Prop 1}$-$\eqref{P Prop 3}, we get
\be\label{P Prop 4}
\hat{q}_i\hat{q}_{m_1} \hat{q}_{m_3} M_{j m_2 m_3 i m_1}(\hat{q})=0\;.
\ee

\section*{Appendix B: Stripping Off Delta Functions}
\renewcommand{\theequation}{B-\Roman{equation}}
\setcounter{equation}{0}

In this Section, we consider the question about how to remove delta functions in correlation functions and write the Ward identities in terms of  on-shell correlators. This was partially done in \cite{Hinterbichler:2013dpa}. We will perform a more systematic analysis here.

Notice that the terms appearing in the original Ward Identities \eqref{rhsfinal} take the general form of 
\begin{align}\label{Abstract Form}
{\cal A}_n=\sum_{b=1}^N \Theta_b(\vec k_b)
\frac{\pd^n}{\pd k_b^{m_1}...\pd k_b^{m_n}}
\left(
	f_b({\vec k}_1,..., {\vec k}_N) \delta(\vec K_t)
\right)\;,
\end{align}
where we denote by $\vec K_t=\vec k_1 +...+\vec k_N$ the {\it total} momentum, and by $\Theta$ collectively the operators such as $\Upsilon, \Gamma,\dots$ appearing in \eqref{rhsfinal}.  The first task is to write ${\cal A}_n$ explicitly in terms of the delta function and its derivatives.

To do that, first define $N-1$ {\it relative} momenta $\vec K_r ^{(i)},\; i=1,...,N-1$. Although the choice of $\vec K_r$ is arbitrary, once it is made all the $N$ momenta $\vec k_b$ --- and hence $\Theta_b(\vec k_b)$ and $f_b({\vec k}_1,..., {\vec k}_N)$ --- can be expressed unambiguously as a function of $\vec K_t$ and $\{\vec K_r\}_{i=1}^{N-1}$: 
\be
\vec k_a=\vec k_a (\vec K_t\;;\vec K_r)\;.
\ee
As a result, ${\cal A}_0$ can be rewritten as 
\be
{\cal A}_0=\sum_{b=1}^N \Theta_b
\big(\vec k_b(0;\vec K_r)\big)f_b(\vec K_t=0; \vec K_r)\delta(\vec K_t)
\equiv \tilde{{\cal A}}_0 \delta(\vec K_t)\;.
\ee

Now let us work out the expression for ${\cal A}_1$: 
\begin{align}
{\cal A}_1&=\sum_{b=1}^N \Theta_b\big(\vec k_b(0;{\vec K}_r)\big)
\left[
\left(\frac{\pd}{\pd k_b^{m_1}}f_b(0;\vec K_r)\right) \delta(\vec K_t)+
f_b(0;\vec K_r)\left(\frac{\pd}{\pd K_t^{m_1}}\delta(\vec K_t)\right) 
\right]
\nonumber\\
&= \delta(\vec K_t)\sum_{b=1}^N \Theta_b(\vec k_b(0;\vec K_r))
\left(\frac{\pd}{\pd k_b^{m_1}}f_b(0;\vec K_r)\right)
+\tilde{\cal A}_0 \frac{\pd}{\pd K_t^{m_1}}\delta(\vec K_t)
\nonumber\\
&+\left(\frac{\pd}{\pd K_t^{m_1}}\delta(\vec K_t)\right) \sum_{b=1}^N f_b(0;\vec K_r)
\left[
	\Theta_b\big( \vec k_b(\vec K_t;\vec K_r) \big) 
	- \Theta_b\big( \vec k_b(0;\vec K_r) \big)
\right]\;.
\end{align}
The last line of the above expression can be further simplified: 
\begin{align}
\text {Last Line} &=\sum_{b=1}^N f_b(0;\vec K_r)
\frac{\pd}{\pd K_t^{m_1}}\left[ 
	\delta(\vec K_t)
	\left(
	\Theta_b\big( \vec k_b(\vec K_t;\vec K_r) \big) 
	- \Theta_b\big( \vec k_b(0;\vec K_r) \big)
	\right)
\right]
\nonumber\\
&-\sum_{b=1}^N f_b(0;\vec K_r)\delta(\vec K_t)\frac{\pd}{\pd K_t^{m_1}}\Theta_b\big( \vec k_b(\vec K_t;\vec K_r) \big)
\nonumber\\
&=-\sum_{b=1}^N f_b(0;\vec K_r)\delta(\vec K_t)\frac{\pd \Theta_b\big( \vec k_b(\vec K_t;\vec K_r) \big)}{\pd K_t^{m_1}}\Bigg\vert_{\vec K_t =0}\;,
\end{align}
where in the last equality we have used the identity about the delta function: $\delta(x)[f(x)-f(0)]=0$. Therefore we get
\begin{align}
{\cal A}_1=\tilde{{\cal A}}_1 \delta(\vec K_t)
+\tilde{{\cal A}}_0 \frac{\pd}{\pd K_t^{m_1}}\delta(\vec K_t)\;,
\end{align}
where we have defined 
\begin{align}
\tilde{{\cal A}}_1=\sum_{b=1}^N 
\left[
\Theta_b(\vec k_b(0;\vec K_r))
\frac{\pd f_b(0;\vec K_r)}{\pd k_b^{m_1}}
-f_b(0;\vec K_r)\frac{\pd \Theta_b\big( \vec k_b(\vec K_t;\vec K_r) \big)}{\pd K_t^{m_1}}\Bigg\vert _{\vec K_t=0}
\right]\;.
\end{align}


Now we are in a position to find the expression for ${\cal A}_n$ for generic $n$. Before we dive into that, let us prove the following lemma about the delta function:

{\bf Lemma} {\it Let $f(x)$ be some smooth function which vanishes sufficiently fast at infinity $|x|\to \infty$. Then
\be\label{n der delta}
f(x) \delta^{(n)}(x)=\sum_{p=0}^{n}(-1)^p \binom{n}{p} f^{(p)}(0)\delta^{(n-p)}(x)\;,
\ee
where $\delta^{(n)}(x)\equiv \frac{{\rm d}^n}{{\rm d} x^n}\delta(x)$ and $f^{(n)}(0)\equiv \frac{{\rm d}^n}{{\rm d} x^n}f(x)\Big\vert_{x=0}$.
}

{\bf Proof} \; For $n=0$, it follows from the property of the delta function. For $n=1$, 
\begin{align}
f(x)\delta'(x)&=\frac{{\rm d}}{{\rm d}x}\left[(f(x)-f(0))\delta(x)\right]
+f(0)\delta'(x)
-f'(x)\delta(x)\nonumber\\
&=f(0)\delta'(x)
-f'(x)\delta(x)\;,
\end{align}
which is precisely the hypothetical equation for $n=1$. 

Now we assume the Eqn.~\eqref{n der delta} holds for all $n\le K-1,\; K\in \mathbb{Z}\cap [1,\infty)$. For $n=K$, 
\begin{align}
f(x)\delta^{(K)}(x)&=\frac{{\rm d}}{{\rm d} x}\left(f(x)\delta^{(K-1)}(x)\right)-f'(x)\delta^{(K-1)}(x)
\nonumber\\
&=\sum_{p=0}^{K-1}(-1)^p \binom{K-1}{p} f^{(p)}(0)\delta^{(K-p)}(x)-
\sum_{p=0}^{K-1}(-1)^p \binom{K-1}{p} f^{(p+1)}(0)\delta^{(K-1-p)}(x) \nonumber\\
&=\sum_{p=0}^{K-1}(-1)^p \binom{K-1}{p} f^{(p)}(0)\delta^{(K-p)}(x)+
\sum_{p=1}^{K}(-1)^p \binom{K-1}{p-1} f^{(p)}(0)\delta^{(K-p)}(x)  \nonumber\\
&=\sum_{p=0}^{K}(-1)^p \binom{K}{p} f^{(p)}(0)\delta^{(K-p)}(x)\;.
\end{align}
Therefore we can conclude that the hypothetical equation \eqref{n der delta} holds for all integers $n$. $\blacksquare$

Now let us go back to ${\cal A}_n$. Using the Leibniz and the above lemma, we get that\footnote{Keep in mind that all the free index $m_1,\dots, m_n$ here are contracted with the symmetric part of the projector $M_{im_0\dots m_n}(\hat q)$.}
\begin{align}
{\cal A}_n
&=\sum_{b=1}^N \Theta_b(\vec k_b)\sum_{p=0}^{n}
\binom{n}{p}
\left[
\frac{\pd^p}{\pd k_b^{m_1}...\pd k_b^{m_p}}  f_b(0;\vec K_r)\right]
\delta^{(n-p)}(\vec K_t)
\nonumber\\
&=\sum_{b=1}^N \sum_{p=0}^{n}
\binom{n}{p}
\frac{\pd^p}{\pd k_b^{m_1}...\pd k_b^{m_p}}f_b(0;\vec K_r)
\sum_{q=0}^{n-p}
(-1)^q \binom{n-p}{q}
\frac{\pd^q  \Theta_b(\vec k_b(\vec K_t;\,\vec K_r))}{\pd K_t^{m_{p+1}}...\pd K_t^{m_{p+q}}}\Bigg\vert_{\vec K_t=0}
\delta^{(n-p-q)}(\vec K_t)\;.
\end{align}
Changing the dummy variable $q$ to $L\equiv q+p$ and noticing that 
\be
\binom{n}{p}\binom{n-p}{q}=\binom{n}{p+q}\binom{p+q}{p}
\ee
 we can rewrite the sum in the equation above as 
\begin{align}\label{An}
{\cal A}_n&=
\sum_{L=0}^{n}
\binom{n}{L}
\delta^{(n-L)}(\vec K_t)
\sum_{b=1}^N 
\sum_{p=0}^L 
(-1)^{L-p} \binom{L}{p}
\frac{\pd^p}{\pd k_b^{m_1}...\pd k_b^{m_p}}f_b(0;\vec K_r)
\frac{\pd^{L-p}  \Theta_b(\vec k_b(\vec K_t;\,\vec K_r))}{\pd K_t^{m_{p+1}}...\pd K_t^{m_{L}}}\Bigg\vert_{\vec K_t=0}
\nonumber\\
&\equiv \sum_{L=0}^{n}
\binom{n}{L}
\delta^{(n-L)}(\vec K_t)
\tilde{\cal A}_L\;,
\end{align}
with 
\be\label{An Expression}
\tilde{\cal A}_L=\sum_{b=1}^N 
\sum_{p=0}^L 
(-1)^{L-p} \binom{L}{p}
\frac{\pd^p}{\pd k_b^{m_1}...\pd k_b^{m_p}}f_b(0;\vec K_r)
\frac{\pd^{L-p}  \Theta_b(\vec k_b(\vec K_t;\,\vec K_r))}{\pd K_t^{m_{p+1}}...\pd K_t^{m_{L}}}\Bigg\vert_{\vec K_t=0}\;.
\ee

The formidable expression for $\tilde{\cal A}_L$ given above can be greatly simplified if we specify the on-shell condition to be 
\be
\vec K_r^{(1)}=\vec k_1\;,\quad \dots\;, \quad \vec K_r^{(N-1)}=\vec k_{N-1}\;,\quad {\vec K}_T=\vec k_1+\dots + \vec k_N\;,
\ee
which can also be written reversely as 
\be
\vec k_1=\vec K_r^{(1)}\;,\quad \dots\;, \quad  \vec k_{N-1}=K_r^{(N-1)}\;,\quad \vec k_N ={\vec K}_T-\vec k_1-\dots - \vec k_{N-1}\;.
\ee
Then it is easy to see that 
\begin{align}
\frac{\pd^\ell }{\pd K_t^{m_1}...\pd K_t^{m_\ell}}\Theta_b\big(\vec k_b(0;\vec K_r)\big)
&=\sum_{m_1', \dots m_\ell'} \frac{\pd k_b^{m_1'}}{\pd K_t^{m_1}}...\frac{\pd k_b^{m_\ell'}}{\pd K_t^{m_\ell}}\frac{\pd^\ell }{\pd k_b^{m_1'}...\pd k_b^{m_\ell'}}\Theta_b(\vec k_b)=0\;, \quad \text{for } \ell\ge 1 \text{ and }b\ne N\;, \nonumber\\
\frac{\pd^\ell}{\pd k_N^{m_1}...\pd k_N^{m_\ell}}f_N(0;\vec K_r)&=
\frac{\pd^\ell }{\pd k_N^{m_1}...\pd k_N^{m_\ell}}f_N(\vec k_1\;,\dots \vec k_{N-1})=0\;,\quad \text{for }\ell \ge 1\;.
\end{align}
Therefore the only terms that survive in $\tilde{A}_L$ are 
\begin{align}\label{ALtilde OS}
\tilde{\cal A}_L&=\sum_{b=1}^{N-1} 
\left[\frac{\pd^L}{\pd k_b^{m_1}...\pd k_b^{m_L}}f_b(0;\vec K_r)\right]
\Theta_b(\vec k_b(0;\,\vec K_r))
+
(-1)^{L}
f_N(0;\vec K_r)
\frac{\pd^{L}  \Theta_N(\vec k_N(\vec K_t;\,\vec K_r))}{\pd K_t^{m_{1}}...\pd K_t^{m_{L}}}\Bigg\vert_{\vec K_t=0}
\nonumber\\
&=\sum_{b=1}^{N-1} 
\left[\frac{\pd^L}{\pd k_b^{m_1}...\pd k_b^{m_L}}f_b(0;\vec K_r) \right]
\Theta_b(\vec k_b(0;\,\vec K_r))
+
(-1)^{L}
f_N(0;\vec K_r)
\frac{\pd^{L}  \Theta_N(\vec k_N)}{\pd k_N^{m_{1}}...\pd k_N^{m_{L}}}\Bigg\vert_{\vec k_N=-\vec k_1-\dots-\vec k_{N-1}}
\;.
\end{align}

\subsection*{B.1 Ward Identities without the Delta Function}
We have mentioned that all the terms in the original Ward identities take the form of \eqref{Abstract Form}, so \eqref{rhsfinal} can be written abstractly as 
\be
I_n=\sum_{i} {\cal A}_n^{(i)}+\frac{1}{n+1}{\cal B}_{n+1} =0\,,
\ee
where we have used ${\cal B}_{n+1}$ unambiguously to denote the term $\sum_{a=1}^N k_a^i \sfrac{\pd^{n+1}}{\pd k_a^{n+1}}\la{\cal O}\ra\delta({\vec K}_t)$ and $i$ to label to other terms in \eqref{rhsfinal}. Using \eqref{An},  we find that 
\begin{align}
0=I_n &= \sum_{L=0}^n \binom{n}{L}\delta^{(n-L)}(\vec K_t) \left[\sum_i\tilde{\cal A}_n^{(i)}\right]
+\frac{1}{n+1}\sum_{L=0}^{n+1}\binom{n+1}{L}\delta^{(n+1-L)}(\vec K_t) \tilde{\cal B}_L \nonumber\\
&=\sum_{L=0}^n \binom{n}{L}\delta^{(n-L)}(\vec K_t) \left[\sum_i\tilde{\cal A}_L^{(i)}+\frac{1}{L+1} \tilde{\cal B}_{L+1} \right]\;,
\end{align}
where in the last equality, we have changed the dummy indice $L$ to $L'=L-1$ and made used the fact that 
\be
\tilde{\cal B}_0=\sum_{a=1}^N k_a^i \la {\cal O}\ra\Bigg\vert_{\vec K_t=0} =0 \;.
\ee
Therefore, the $n=0$ Ward Identity is equivalent to \footnote{Notice that $\tilde{\cal A}_L$ and $\tilde{\cal B}_L$ are already computed on shell.} $\sum_i\tilde{\cal A}_0^{(i)}+\tilde{\cal B}_{1}=0$. For $n\ge 1$, the new information contained in each higher order identity is 
\be\label{WI delta remove}
\sum_i\tilde{\cal A}_n^{(i)}+\frac{1}{n+1}\tilde{\cal B}_{n+1}=0\;.
\ee
Since the delta function has been removed from $\tilde{\cal A}_n$ and $\tilde{\cal B}_n$, \eqref{WI delta remove} is in the form that we are after. 

It remains to work out explicitly each term in the Ward identities in this delta-function-removed form, by applying \eqref{ALtilde OS}.  For instance, for the term on the left-hand side of \eqref{rhsfinal}, effectively we can think it as having $\Theta_q(\vec q)=1\;, \Theta_b(\vec k_b)=0$ for $b=1,\dots N$. Therefore, omitting indices of the correlation functions, we have
\begin{align}
\text{L.~H.~S.~} & \to
\frac{\pd^n}{\pd q^{m_1}...\pd q^{m_n}}
\left(
\frac{\la \gamma(\vec q) {\cal O}\ra}{\la \gamma \gamma\ra}\Bigg\vert_{\vec q+\vec K_t=0}
+\frac{\la \zeta(\vec q) {\cal O}\ra}{3 \la\zeta \zeta \ra}\Bigg\vert_{\vec q+\vec K_t=0}
\right)
\nonumber\\
&=
\frac{\pd^n}{\pd q^{m_1}...\pd q^{m_n}}
\left(
\frac{\la \gamma(\vec q) {\cal O}\ra'}{P_\gamma(q)}
+\frac{\la \zeta(\vec q) {\cal O}\ra'}{3 P_\zeta(q)}
\right)
\;,
\end{align}
where we have denoted by $\la\dots \ra'$ the correlation function with delta function removed and with on-shell condition imposed. We can play a similar trick on the terms on the right hand side of \eqref{rhsfinal}. Finally we reach the Ward identities expressed in terms of primed correlators: 

\underline{For ${\cal O}$ containing at least one $\gamma$ field: }
\bea
\nonumber
 & & \lim_{\vec{q}\rightarrow 0} M_{i\ell_0 \ldots \ell_n}(\hat{q}) \frac{\partial^{n}}{\partial q_{\ell_1}\cdots \partial q_{\ell_n}} \Bigg(\frac{1}{P_\gamma(q)} \la \gamma^{i\ell_0}(\vec{q}){\cal O}(\vec{k}_1,\ldots,\vec{k}_N) \ra_c' + \frac{\delta^{i\ell_0}}{3P_\zeta(q)} \la  \zeta(\vec{q}) {\cal O}(\vec{k}_1,\ldots,\vec{k}_N) \ra_c'  \Bigg) \\
\nonumber
& & ~~ =  - M_{i\ell_0 \ldots \ell_n}(\hat{q}) \Bigg\{ \sum_{a=1}^{N-1} \Bigg( \delta^{i\ell_0} \frac{\partial^{n}}{\partial k_{\ell_1}^a\cdots \partial k_{\ell_n}^a} - \frac{\delta_{n0}}{N-1}\delta^{i\ell_0}
+ \frac{k^{i}_a}{n+1}  \frac{\partial^{n+1}}{\partial k_{\ell_0}^a \cdots \partial k_{\ell_n}^a}\Bigg) \la  {\cal O}(\vec{k}_1,\ldots,\vec{k}_N) \ra_c'  \\
\nonumber
& & \;\;\;\;~~-\sum_{a=1}^M \Upsilon^{i\ell_0i_aj_a}(\hat{k}_a) \Big\vert_{\rm OS}\; \frac{\partial^{n}}{\partial k_{\ell_1}^a\cdots \partial k_{\ell_n}^a} \la {\cal O}^\zeta(\vec{k}_1,\ldots,\vec{k}_{a-1},\vec{k}_{a+1},\ldots \vec{k}_M)\gamma_{i_aj_a}(\vec{k}_a) {\cal O}^\gamma (\vec{k}_{M+1},\ldots,\vec{k}_N)\ra_c' \\
\nonumber
& & \;\;\;\; ~~ -\sum_{b=M+1}^{N-1}  \Gamma^{i\ell_0\;\;\;\;\;k_b\ell_b}_{\;\;\;\;i_bj_b}(\hat{k}_b )\Big\vert_{\rm OS}\;\frac{\partial^{n}}{\partial k_{\ell_1}^b\cdots \partial k_{\ell_n}^b}\la {\cal O}^\zeta(\vec{k}_1,\ldots,\vec{k}_M) {\cal O}^\gamma_{i_{M+1} j_{M+1},\ldots,k_b\ell_b,\ldots i_Nj_N}(\vec{k}_{M+1},\ldots,\vec{k}_N) \ra_c' \\ 
\nonumber
& & \;\;\;\; ~~ -(-1)^n
\left[\frac{\partial^{n}}{\partial k_{\ell_1}^N\cdots \partial k_{\ell_n}^N}\Gamma^{i\ell_0\;\;\;\;\;k_N\ell_N}_{\;\;\;\;i_N j_N}(\hat{k}_N )
\right]\Bigg\vert_{\rm OS}\nonumber \\
& &\;\;\;\;~~\times \la {\cal O}^\zeta(\vec{k}_1,\ldots,\vec{k}_M) {\cal O}^\gamma_{i_{M+1} j_{M+1},\ldots i_{N-1}j_{N-1},k_N\ell_N,}(\vec{k}_{M+1},\ldots,\vec{k}_N) \ra_c' 
\Bigg\} \nonumber\\
& &  \;\;\;\; ~~+ \ldots \;,
\eea

\underline{For ${\cal O}={\cal O}^\zeta$:}
\bea
\nonumber
 & & \lim_{\vec{q}\rightarrow 0} M_{i\ell_0 \ldots \ell_n}(\hat{q}) \frac{\partial^{n}}{\partial q_{\ell_1}\cdots \partial q_{\ell_n}} \Bigg(\frac{1}{P_\gamma(q)} \la \gamma^{i\ell_0}(\vec{q}){\cal O}(\vec{k}_1,\ldots,\vec{k}_N) \ra_c' + \frac{\delta^{i\ell_0}}{3P_\zeta(q)} \la  \zeta(\vec{q}) {\cal O}(\vec{k}_1,\ldots,\vec{k}_N) \ra_c'  \Bigg) \\
\nonumber
& & ~~ =  - M_{i\ell_0 \ldots \ell_n}(\hat{q}) \Bigg\{ \sum_{a=1}^{N-1} \Bigg( \delta^{i\ell_0} \frac{\partial^{n}}{\partial k_{\ell_1}^a\cdots \partial k_{\ell_n}^a} - \frac{\delta_{n0}}{N-1}\delta^{i\ell_0}
+ \frac{k^{i}_a}{n+1}  \frac{\partial^{n+1}}{\partial k_{\ell_0}^a \cdots \partial k_{\ell_n}^a}\Bigg) \la  {\cal O}(\vec{k}_1,\ldots,\vec{k}_N) \ra_c'  \\
\nonumber
& & \;\;\;\;~~-\sum_{a=1}^{N-1} \Upsilon^{i\ell_0i_aj_a}(\hat{k}_a) \Big\vert_{\rm OS}\; \frac{\partial^{n}}{\partial k_{\ell_1}^a\cdots \partial k_{\ell_n}^a} \la {\cal O}^\zeta(\vec{k}_1,\ldots,\vec{k}_{a-1},\vec{k}_{a+1},\ldots \vec{k}_N)\gamma_{i_aj_a}(\vec{k}_a) \ra_c'\nonumber\\ 
& & \;\;\;\;~~-(-1)^n 
\left[\frac{\partial^{n}}{\partial k_{\ell_1}^N\cdots \partial k_{\ell_n}^N}\Upsilon^{i\ell_0i_Nj_N}(\hat{k}_N)\right] \Bigg\vert_{\rm OS}\;  
\la {\cal O}^\zeta(\vec{k}_1,\ldots,\vec{k}_{N-1})\gamma_{i_N j_N}(\vec{k}_N)\ra_c'
\Bigg\}
\nonumber \\
& &  \;\;\;\; ~~+ \ldots \;,
\eea

where the on-shell condition is specified explicitly by $\vec k_N = -\vec k_1 -\dots -\vec k_{N-1}$\;.

\bibliographystyle{utphys}
\addcontentsline{toc}{section}{References}
\bibliography{consistencychecks9}

\providecommand{\href}[2]{#2}\begingroup\raggedright\begin{thebibliography}{10}

\bibitem{Maldacena:2002vr}
J.~M. Maldacena, ``{Non-Gaussian features of primordial fluctuations in single
  field inflationary models},''
  \href{http://dx.doi.org/10.1088/1126-6708/2003/05/013}{{\em JHEP} {\bf 0305}
  (2003)  013},
\href{http://arxiv.org/abs/astro-ph/0210603}{{\tt arXiv:astro-ph/0210603
  [astro-ph]}}.

\bibitem{Creminelli:2004yq}
P.~Creminelli and M.~Zaldarriaga, ``{Single field consistency relation for the
  3-point function},''
  \href{http://dx.doi.org/10.1088/1475-7516/2004/10/006}{{\em JCAP} {\bf 0410}
  (2004)  006},
\href{http://arxiv.org/abs/astro-ph/0407059}{{\tt arXiv:astro-ph/0407059
  [astro-ph]}}.

\bibitem{Cheung:2007sv}
C.~Cheung, A.~L. Fitzpatrick, J.~Kaplan, and L.~Senatore, ``{On the consistency
  relation of the 3-point function in single field inflation},''
  \href{http://dx.doi.org/10.1088/1475-7516/2008/02/021}{{\em JCAP} {\bf 0802}
  (2008)  021},
\href{http://arxiv.org/abs/0709.0295}{{\tt arXiv:0709.0295 [hep-th]}}.

\bibitem{Creminelli:2012ed}
P.~Creminelli, J.~Norena, and M.~Simonovic, ``{Conformal consistency relations
  for single-field inflation},''
  \href{http://dx.doi.org/10.1088/1475-7516/2012/07/052}{{\em JCAP} {\bf 1207}
  (2012)  052},
\href{http://arxiv.org/abs/1203.4595}{{\tt arXiv:1203.4595 [hep-th]}}.

\bibitem{Hinterbichler:2012nm}
K.~Hinterbichler, L.~Hui, and J.~Khoury, ``{Conformal Symmetries of Adiabatic
  Modes in Cosmology},''
  \href{http://dx.doi.org/10.1088/1475-7516/2012/08/017}{{\em JCAP} {\bf 1208}
  (2012)  017},
\href{http://arxiv.org/abs/1203.6351}{{\tt arXiv:1203.6351 [hep-th]}}.

\bibitem{Senatore:2012wy}
L.~Senatore and M.~Zaldarriaga, ``{A Note on the Consistency Condition of
  Primordial Fluctuations},''
  \href{http://dx.doi.org/10.1088/1475-7516/2012/08/001}{{\em JCAP} {\bf 1208}
  (2012)  001},
\href{http://arxiv.org/abs/1203.6884}{{\tt arXiv:1203.6884 [astro-ph.CO]}}.

\bibitem{Hinterbichler:2013dpa}
K.~Hinterbichler, L.~Hui, and J.~Khoury, ``{An Infinite Set of Ward Identities
  for Adiabatic Modes in Cosmology},''
\href{http://arxiv.org/abs/1304.5527}{{\tt arXiv:1304.5527 [hep-th]}}.

\bibitem{Goldberger:2013rsa}
W.~D. Goldberger, L.~Hui, and A.~Nicolis, ``{One-particle-irreducible
  consistency relations for cosmological perturbations},''
  \href{http://dx.doi.org/10.1103/PhysRevD.87.103520}{{\em Phys.Rev.} {\bf D87}
  (2013)  103520},
\href{http://arxiv.org/abs/1303.1193}{{\tt arXiv:1303.1193 [hep-th]}}.

\bibitem{Berezhiani:2013ewa}
L.~Berezhiani and J.~Khoury, ``{Slavnov-Taylor Identities for Primordial
  Perturbations},''
\href{http://arxiv.org/abs/1309.4461}{{\tt arXiv:1309.4461 [hep-th]}}.

\bibitem{Pimentel:2013gza}
G.~L. Pimentel, ``{Inflationary Consistency Conditions from a Wavefunctional
  Perspective},''
\href{http://arxiv.org/abs/1309.1793}{{\tt arXiv:1309.1793 [hep-th]}}.

\bibitem{Cai:2009fn}
Y.-F. Cai, W.~Xue, R.~Brandenberger, and X.~Zhang, ``{Non-Gaussianity in a
  Matter Bounce},'' \href{http://dx.doi.org/10.1088/1475-7516/2009/05/011}{{\em
  JCAP} {\bf 0905} (2009)  011},
\href{http://arxiv.org/abs/0903.0631}{{\tt arXiv:0903.0631 [astro-ph.CO]}}.

\bibitem{Khoury:2008wj}
J.~Khoury and F.~Piazza, ``{Rapidly-Varying Speed of Sound, Scale Invariance
  and Non-Gaussian Signatures},''
  \href{http://dx.doi.org/10.1088/1475-7516/2009/07/026}{{\em JCAP} {\bf 0907}
  (2009)  026},
\href{http://arxiv.org/abs/0811.3633}{{\tt arXiv:0811.3633 [hep-th]}}.

\bibitem{Namjoo:2012aa}
M.~H. Namjoo, H.~Firouzjahi, and M.~Sasaki, ``{Violation of non-Gaussianity
  consistency relation in a single field inflationary model},''
  \href{http://dx.doi.org/10.1209/0295-5075/101/39001}{{\em Europhys.Lett.}
  {\bf 101} (2013)  39001},
\href{http://arxiv.org/abs/1210.3692}{{\tt arXiv:1210.3692 [astro-ph.CO]}}.

\bibitem{Chen:2013aj}
X.~Chen, H.~Firouzjahi, M.~H. Namjoo, and M.~Sasaki, ``{A Single Field
  Inflation Model with Large Local Non-Gaussianity},''
  \href{http://dx.doi.org/10.1209/0295-5075/102/59001}{{\em Europhys.Lett.}
  {\bf 102} (2013)  59001},
\href{http://arxiv.org/abs/1301.5699}{{\tt arXiv:1301.5699 [hep-th]}}.

\bibitem{Chen:2013eea}
X.~Chen, H.~Firouzjahi, E.~Komatsu, M.~H. Namjoo, and M.~Sasaki, ``{In-in and
  $\delta N$ calculations of the bispectrum from non-attractor single-field
  inflation},'' \href{http://dx.doi.org/10.1088/1475-7516/2013/12/039}{{\em
  JCAP} {\bf 1312} (2013)  039},
\href{http://arxiv.org/abs/1308.5341}{{\tt arXiv:1308.5341 [astro-ph.CO]}}.

\bibitem{Kehagias:2013yd}
A.~Kehagias and A.~Riotto, ``{Symmetries and Consistency Relations in the Large
  Scale Structure of the Universe},''
  \href{http://dx.doi.org/10.1016/j.nuclphysb.2013.05.009}{{\em Nucl.Phys.}
  {\bf B873} (2013)  514--529},
\href{http://arxiv.org/abs/1302.0130}{{\tt arXiv:1302.0130 [astro-ph.CO]}}.

\bibitem{Peloso:2013zw}
M.~Peloso and M.~Pietroni, ``{Galilean invariance and the consistency relation
  for the nonlinear squeezed bispectrum of large scale structure},''
  \href{http://dx.doi.org/10.1088/1475-7516/2013/05/031}{{\em JCAP} {\bf 1305}
  (2013)  031},
\href{http://arxiv.org/abs/1302.0223}{{\tt arXiv:1302.0223 [astro-ph.CO]}}.

\bibitem{Creminelli:2013mca}
P.~Creminelli, J.~Nore–a, M.~Simonovic, and F.~Vernizzi, ``{Single-Field
  Consistency Relations of Large Scale Structure},''
\href{http://arxiv.org/abs/1309.3557}{{\tt arXiv:1309.3557 [astro-ph.CO]}}.

\bibitem{Creminelli:2013poa}
P.~Creminelli, J.~Gleyzes, M.~Simonovic, and F.~Vernizzi, ``{Single-Field
  Consistency Relations of Large Scale Structure. Part II: Resummation and
  Redshift Space},''
\href{http://arxiv.org/abs/1311.0290}{{\tt arXiv:1311.0290 [astro-ph.CO]}}.

\bibitem{Creminelli:2013nua}
P.~Creminelli, J.~Gleyzes, L.~Hui, M.~Simonovic, and F.~Vernizzi,
  ``{Single-Field Consistency Relations of Large Scale Structure. Part III:
  Test of the Equivalence Principle},''
\href{http://arxiv.org/abs/1312.6074}{{\tt arXiv:1312.6074 [astro-ph.CO]}}.

\bibitem{Peloso:2013spa}
M.~Peloso and M.~Pietroni, ``{Ward identities and consistency relations for the
  large scale structure with multiple species},''
\href{http://arxiv.org/abs/1310.7915}{{\tt arXiv:1310.7915 [astro-ph.CO]}}.

\bibitem{Kehagias:2013rpa}
A.~Kehagias, J.~Nore–a, H.~Perrier, and A.~Riotto, ``{Consequences of
  Symmetries and Consistency Relations in the Large-Scale Structure of the
  Universe for Non-local bias and Modified Gravity},''
\href{http://arxiv.org/abs/1311.0786}{{\tt arXiv:1311.0786 [astro-ph.CO]}}.

\bibitem{Valageas:2013cma}
P.~Valageas, ``{Consistency relations of large-scale structures},''
\href{http://arxiv.org/abs/1311.1236}{{\tt arXiv:1311.1236 [astro-ph.CO]}}.

\bibitem{Assassi:2012zq}
V.~Assassi, D.~Baumann, and D.~Green, ``{On Soft Limits of Inflationary
  Correlation Functions},''
  \href{http://dx.doi.org/10.1088/1475-7516/2012/11/047}{{\em JCAP} {\bf 1211}
  (2012)  047},
\href{http://arxiv.org/abs/1204.4207}{{\tt arXiv:1204.4207 [hep-th]}}.

\bibitem{Weinberg:2003sw}
S.~Weinberg, ``{Adiabatic modes in cosmology},''
  \href{http://dx.doi.org/10.1103/PhysRevD.67.123504}{{\em Phys.Rev.} {\bf D67}
  (2003)  123504},
\href{http://arxiv.org/abs/astro-ph/0302326}{{\tt arXiv:astro-ph/0302326
  [astro-ph]}}.

\bibitem{Creminelli:2012xb}
P.~Creminelli, J.~Norena, M.~Pena, and M.~Simonovic, ``{Khronon inflation},''
  \href{http://dx.doi.org/10.1088/1475-7516/2012/11/032}{{\em JCAP} {\bf 1211}
  (2012)  032},
\href{http://arxiv.org/abs/1206.1083}{{\tt arXiv:1206.1083 [hep-th]}}.

\bibitem{Cheung:2007st}
C.~Cheung, P.~Creminelli, A.~L. Fitzpatrick, J.~Kaplan, and L.~Senatore, ``{The
  Effective Field Theory of Inflation},''
  \href{http://dx.doi.org/10.1088/1126-6708/2008/03/014}{{\em JHEP} {\bf 0803}
  (2008)  014},
\href{http://arxiv.org/abs/0709.0293}{{\tt arXiv:0709.0293 [hep-th]}}.

\bibitem{Creminelli:2013cga}
P.~Creminelli, A.~Perko, L.~Senatore, M.~Simonovic, and G.~Trevisan, ``{The
  Physical Squeezed Limit: Consistency Relations at Order $q^2$},''
  \href{http://dx.doi.org/10.1088/1475-7516/2013/11/015}{{\em JCAP} {\bf 1311}
  (2013)  015},
\href{http://arxiv.org/abs/1307.0503}{{\tt arXiv:1307.0503 [astro-ph.CO]}}.

\bibitem{Maldacena:2011nz}
J.~M. Maldacena and G.~L. Pimentel, ``{On graviton non-Gaussianities during
  inflation},'' \href{http://dx.doi.org/10.1007/JHEP09(2011)045}{{\em JHEP}
  {\bf 1109} (2011)  045},
\href{http://arxiv.org/abs/1104.2846}{{\tt arXiv:1104.2846 [hep-th]}}.

\bibitem{usfuture}
L.~Berezhiani, J.~Khoury, and J.~Wang, ``{Violations of Novel Inflationary
  Consistency Relations},'' {\em to appear}  .

\bibitem{Weinberg:1966kf}
S.~Weinberg, ``{Pion scattering lengths},''
\href{http://dx.doi.org/10.1103/PhysRevLett.17.616}{{\em Phys.Rev.Lett.} {\bf
  17} (1966)  616--621}.

\bibitem{Marcofuture}
A.~Joyce, J.~Khoury, and M.~Simonovic, ``{Multiple Soft Limits of Primordial
  Correlation Functions},'' {\em to appear}  .

\end{thebibliography}\endgroup

\end{document}